\documentclass[aps,prd,showpacs,nofootinbib,tightenlines]{revtex4}
\usepackage{amsmath}
\usepackage{amssymb}
\usepackage{epsfig}
\usepackage{graphicx}
\usepackage{float}
\usepackage{color} %% {\color{red} xxxx }
\usepackage{ulem}  %%\sout{word}
\begin{document}

\title{Virtual contributions from $D^{\ast}(2007)^0$ and $D^{\ast}(2010)^{\pm}$ in the $B\to D\pi h$ decays}
\author{Wen-Fei Wang$^{1,2}$}\email{wfwang@sxu.edu.cn}
\author{Jian Chai$^{1,2}$}
\affiliation{$^1$Institute of Theoretical Physics, Shanxi University, Taiyuan, Shanxi 030006, China}
\affiliation{$^2$State Key Laboratory of Quantum Optics and Quantum Optics Devices, Shanxi University, 
                        Taiyuan, Shanxi 030006, China}

\date{\today}

%XXXXXXXXXXXXXXXXXXXXXXXXXXXXXXXXXX%
\begin{abstract}
  We study the quasi-two-body decays $B\to D^*h \to D\pi h$ with $h=(\pi, K)$ in the perturbative QCD approach
  and focus on the virtual contributions from the off-shell $D^{\ast}(2007)^0$ and $D^{\ast}(2010)^{\pm}$ in the four 
  measured decays $\bar B^0 \to D^0\pi^+\pi^-$, $\bar B^0\to D^0\pi^+K^-$, 
  $B^- \to D^+\pi^-\pi^-$ and $B^- \to D^+\pi^-K^-$.
  For the $\bar B^0 \to D^{*+}\pi^-\to D^0\pi^+\pi^-$ and $\bar B^0\to D^{*+}K^-\to D^0\pi^+K^-$ decays, 
  their branching fractions concentrate in a very small region of $m_{D^0\pi^+}$ near  $D^{*+}$ pole mass, 
  and the virtual contributions from $D^{*+}$, in the region $m_{D^0\pi^+}>2.1$ GeV, are about $5\%$ of the corresponding 
  quasi-two-body results. We define two ratios $R_{D^{*+}}$ and $R_{D^{*0}}$, from which we conclude that the flavor-$SU(3)$ 
  symmetry will be maintained for the $B\to D^* h\to D\pi h$ decays with very small breaking at any physical value of the $m_{D\pi}$.
  The $B^-\to D^{*0}\pi^-\to D^0\pi^0\pi^-$ and $B^-\to D^{*0}K^-\to D^0\pi^0K^-$ decays can be employed as a constraint 
  for the $D^{*0}$ decay width, with preferred values consistent with previous theoretical predictions for this quantity.
\end{abstract}

\pacs{13.20.He, 13.25.Hw, 13.30.Eg}
\maketitle
%XXXXXXXXXXXXXXXXXXXXXXXXXXXXXXXXXX% @ Begin

\section{INTRODUCTION}
Three-body hadronic decays $B\to D\pi h$, with the $h$ is pion or kaon, have been suggested as a way to measure the
Cabibbo-Kobayashi-Maskawa (CKM)~\cite{CKM-C,CKM-KM} angle $\gamma$~\cite{prd67-096002,prd79-051301,
prd80-092002,prd81-014025,prd97-056002} (which have been performed in~\cite{prd78-034023,prd93-112018,jhep1612-087}) 
and angle $\beta$~\cite{plb425-375,jpg36-025006,jhep1803-195}. These decay processes have also been proven as 
an appropriate field for the studies of charm meson spectroscopy, the Belle, BaBar and LHCb Collaborations have 
achieved brilliant progress in identifying the excited charm states and measuring their parameters~\cite{prd69-112002,
prd76-012006,prd79-112004,prd91-092002,prd92-012012,prd92-032002,prd93-051101,prd94-072001}.
In the amplitude analyses of $B\to D\pi h$ decays, one has the contributions from the quasi-two-body decay processes
$B\to D^*h\to D\pi h$, including the $S$-wave ground states of the $c\bar q$ ($q$ is $u$ or $d$) quark system, the 
charmed vectors $D^{\ast}(2007)^0$ and $D^{\ast}(2010)^\pm$~\cite{Goldhaber,Nguyen1977,Peruzzi} as the intermediate states. 
With the strong kinematic suppression, the charged state $D^{\ast+}$ may decay into $D^0\pi^+$ or $D^+\pi^0$, the neutral state 
$D^{\ast 0}$ can decay into $D^0\pi^0$. The natural decay mode $D^{\ast 0}\to D^+\pi^-$ for the $D^{\ast 0}$ is blocked
because of its pole mass and the threshold of its decay daughters. 

The $D^*$ is usually studied, on the theoretical side, as the stable particle in two-body hadronic $B$ meson decays in the literature. 
The discussions of the factorization formula for the $B$ meson decays to $D^{(*)}$ and a light pseudoscalar or vector meson 
could be found in Refs.~\cite{npb591-313,jhep1609-112}. In~\cite{zpc29-637}, the color-favored decays $B\to D^{(*)}\pi$ 
were explored within the factorization hypothesis. Using the factorization approach, the two-body decays $B\to D^* h$ have 
been studied in~\cite{zpc34-103,plb318-549,ijmpa24-5845}. Phenomenological studies of the 
$\bar{B}_{d,s} \to D^{*}_{d,s} V$ decays were performed in~\cite{epjc76-523} within the framework of QCD Factorization. The global 
fits under the assumption of flavor $SU(3)$ symmetry for the charmed $B$ decays have their results in~\cite{prd75-074021}. 
Within the factorization-assisted topological-amplitude approach, the two-body decays $B\to D^{(*)}M$ have been studied 
in~\cite{prd92-094016}. The discussions of the isospin relations for the $B\to D^*K(\pi)$ could be found in Ref.~\cite{prd67-074013}. 
While in the perturbative QCD (PQCD) approach~\cite{plb504-6,prd63-054008,prd63-074009,ppnp51-85}, 
the $B\to D^{(*)}$ form factors and the $B\to D^{(*)}M$ decays were calculated in~\cite{prd67-054028} and~\cite{prd69-094018}, 
respectively, the color suppressed decay modes $B^0\to \bar D^{(*)0} \eta^{(\prime)}$ were analyzed in~\cite{prd68-097502}, 
the two-body $B_{(s)}\to D^{(*)}_{(s)}P$ and $B_{(s)}\to D^{(*)}_{(s)}V$ decays were studied in Ref.~\cite{prd78-014018} and 
$B$ meson decay into $D^{(*)}$ and a light scalar meson were studied in Ref.~\cite{prd95-016011}. 
 
In the Dalitz plot~\cite{dalitz} analyses of the decays $B^-\to D^+\pi^- h^-$ (the inclusion of charge-conjugate processes is always 
implied) performed by Belle~\cite{prd69-112002}, BaBar~\cite{prd79-112004} and LHCb Collaborations~\cite{prd91-092002,
prd94-072001}, the virtual contributions for the $D^+\pi^-$ pair from the intermediate state $D^{*0}$ 
were found to be indispensable for the total amplitudes. 
The virtual contributions are the contributions from the state $D^{\ast 0}$ for the $D^+\pi^-$ pair 
in the quasi-two-body processes $B^-\to D^{*0}\pi^- \to D^+\pi^-\pi^-$ and $B^-\to D^{*0}K^- \to D^+\pi^-K^-$
with the resonance pole mass outside the kinematically accessible region of the phase space~\cite{prd91-092002,prd94-072001}.
That is to say, although the pole mass of $D^{\ast 0}$ is lower than the threshold of $D^+\pi^-$ pair, 
the natural decay tunnel $D^{\ast 0}\to D^+\pi^-$
is blocked, but the resonance tail will contribute to the total branching fractions of the $B^-\to D^+\pi^- h^-$ processes, and the 
off-shell effects were found surprisingly large in~\cite{1806-09853}. For the decays $\bar B^0 \to D^0\pi^+h^-$, the portion of 
$\bar B^0 \to D^{*+}h^-$ with the natural decay $D^{*+}\to D^0\pi^+$ were always excluded from the total three-body branching 
fractions by a cut of the $D^0\pi^+$ invariant mass, while the necessary off-shell effects were retained in the decay 
amplitudes~\cite{prd76-012006,prd92-012012,prd92-032002}.

In order to extract the most information on the involved strong and weak dynamics from the experimental data of the three-body 
$B$ decays, different methods have been adopted, such as the isospin, U-spin and/or flavor $SU(3)$ 
symmetries~\cite{plb564-90,prd72-075013,prd72-094031,prd84-056002,plb727-136,plb726-337,prd89-074043,plb728-579,
ijmpa29-1450011,prd91-014029}, the QCD factorization~\cite{plb622-207,prd74-114009,prd79-094005,prd81-094033,appb42-2013,
prd66-054015,prd72-094003,prd76-094006,prd88-114014,prd89-074025,prd89-094007,epjc75-536,prd87-076007,epjc78-845,
npb899-247,prd96-113003} and the PQCD approach~\cite{plb561-258,prd70-054006,prd89-074031,prd91-094024,
prd97-034033,1803-02656,1810-12507} in abundant works. While 
% because of entangled resonant and nonresonant contributions, and the final-state 
%interactions~\cite{1512-09284,plb780-357}, the strong dynamics contained in three-body $B$ meson decays is 
%much more complicated than that in the two-body cases, together with the complex interplay between the weak 
%processes and the low-energy strong interactions~\cite{epjc77-561}, the traditional approaches techniques and tools 
%for the two-body decays are no longer satisfactory in the three-body processes~\cite{1605-03889}. 
three-body hadronic $B$ decays are known experimentally, in most cases, to be dominated by the low energy scalar, vector 
and tensor resonances, which could be analysed in the quasi-two-body framework by neglecting the three-body effects
and the rescattering effects~\cite{1512-09284,plb763-29}. In the quasi-two-body framework, we always assume two final states, 
in the three-body processes, form a single resonant state which originated from a quark-antiquark pair and then the factorization 
procedure can be applied~\cite{1605-03889,prd96-113003}.
In this work, we will focus on the virtual contributions originated from off-shell $D^*$ in the measured decays 
$B^- \to D^+\pi^-\pi^-$, $B^- \to D^+\pi^-K^-$, $\bar B^0 \to D^0\pi^+\pi^-$ and $\bar B^0 \to D^0\pi^+K^-$. 
The $D^*$ off-shell effects in the four decay processes  $B\to D\pi h$ and the natural contributions $D^{*+}\to D^0\pi^+$ in the two 
$\bar B^0$ decays in this work shall be analysed in the quasi-two-body framework which has been detailed discussed 
in Ref~\cite{plb763-29} in PQCD approach. The method used in~\cite{plb763-29} has been adopted in Refs.~\cite{paps-li-ya,
paps-li-ya-II,paps-ma-aj,1809-02943} for the studies of some quasi-two-body $B$ meson decays.

This work is organized as follows. In Sec.~II, we give a brief introduction for the theoretical framework.
In Sec.~III, we show the numerical results. Discussions and conclusions are given in Sec.~IV.
The factorization formulas for the relevant quasi-two-body decay amplitudes are collected in the Appendix.

\section{FRAMEWORK} %XXXXXXXXXXXX-----<<<FRAMEWORK>>>-----

\begin{figure}[tbp]
%\vspace{-1cm}
\centerline{\epsfxsize=10cm \epsffile{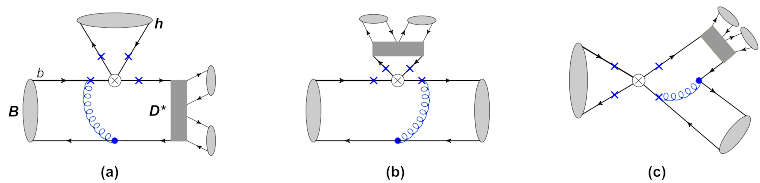}}
\caption{Typical Feynman diagrams for the decay processes $B\to D^* h\to D\pi h$, $h=$($\pi, K$). The symbol $\otimes$ 
is the weak vertex, $\times$ denotes possible attachments of hard gluons and the rectangle represents the vector states $D^*$.}
\label{fig-feyndiag}
\end{figure}

In the rest frame of $B$ meson, with $m_B$ being its mass, we define momentum $p_B$ and the light 
spectator quark momentum $k_B$ for it as
\begin{eqnarray}
p_B=\frac{m_B}{\sqrt2}(1,1,0_{\rm T}),\quad  k_B=\left(0,\frac{m_B}{\sqrt2}x_B ,k_{B{\rm T}}\right),
\end{eqnarray}
in the light-cone coordinates, where $x_B$ is the momentum fraction. 
The momenta $p_3$ and $k_3$ for the bachelor final state $h$ and its spectator quark have their definitions as
\begin{eqnarray}
p_3=\frac{m_B}{\sqrt2}(0,1-\eta,0_{\rm T}),\quad
k_3=\left(0,\frac{m_B}{\sqrt2}(1-\eta)x_3,k_{3{\rm T}}\right).
\end{eqnarray}
For the state $D^*$ and the $D\pi$ pair decays from it in the Feynman diagrams, the Fig.~\ref{fig-feyndiag}, 
for the quasi-two-body processes $B\to D^* h\to D\pi h$, we define their momentum $p=\frac{m_B}{\sqrt 2}(1, \eta,0)$. 
Its easy to see $\eta=s/m_B^2$, with the invariant mass square $s=p^2$ for the $D\pi$ pair. The light spectator quark comes 
from $B$ meson and goes into intermediate state in the $D^*$ hadronization as shown in Fig.~\ref{fig-feyndiag}~(a) has the 
momentum $k=(\frac{m_B}{\sqrt 2}z, 0, k_{\rm T})$. Where $x_3$ and $z$ are the corresponding momentum fractions and 
run from 0 to 1.

%%%%%%%%%%%%%%%%%%%%%%%%%%%%%%%%%%%%%%%%%%%%%%%%%%
%%%%%%%%%%%%%%%%%%%%%%%%%%%%%%%%%%%%%%%%%%%%%%%%%%

The distribution amplitudes for the $B$ meson and the bachelor final state pion or kaon in this work are the same 
as those widely adopted in the PQCD approach in the hadronic B meson decays, one can find their expressions and the 
relevant parameters in Ref.~\cite{prd86-114025}. For the longitudinal polarization structure of the $P$-wave $D\pi$ system 
which including the $D^*$ hadronization and the $D^*\to D\pi$ processes, based on the discussions in 
Refs.~\cite{prd67-054028,prd78-014018,plb763-29,prd91-094024,epjc78-76,1807-03453}, one could write
\begin{eqnarray}
\Phi^{P\text{-wave}}_{D\pi}=\frac{1}{\sqrt{2N_c}}{\epsilon\hspace{-1.6truemm}/}_L\left({p\hspace{-1.6truemm}/ } 
                                             +\sqrt s\right) \phi_{D\pi}(z,b,s), 
\end{eqnarray}
with the distribution amplitude %$\phi_{D\pi}(z,b,s)$
\begin{eqnarray}
\phi_{D\pi}(z,b,s)=\frac{F_{D\pi}(s) }{2\sqrt{2N_c}} 6z(1-z)\left[1+a_{D\pi}(1-2z) \right] {\rm exp}\left(-\omega^2_{D\pi}b^2/2 \right),
\label{eqn-def-DA}
\end{eqnarray}
where the $a_{D\pi}$ and $\omega_{D\pi}$ are the Gegenbauer moment and the shape parameter for the $P$-wave $D\pi$ system, 
respectively. The time-like form factor $F_{D\pi}(s)$ has its definition in the matrix elements
\begin{eqnarray}
\langle D(p_1)\pi(p_2) | \bar c\gamma_\mu(1-\gamma_5) q| 0 \rangle
                               =\left[(p_1-p_2)_\mu - \frac{m^2_D-m^2_\pi}{p^2}p_\mu \right]F_{D\pi}(s)
                               + \frac{m^2_D-m^2_\pi}{p^2}p_\mu F_{0}(s)\;,
\label{eqn-def-ff}                   
\end{eqnarray}  % Eq.~(\ref{eqn-def-ff}), 
where $p=(p_1+p_2)$, $p_1$($p_2$) is the momentum for $D$($\pi$) and $m_D$($m_\pi$) is the mass of $D$($\pi$) meson.
The $F_{0}(s)$ is the $S$-wave form factor for $D\pi$ system. With Eq.~(\ref{eqn-def-ff}), by inserting the intermediate state 
$D^*$, it's easy to have the following expression for the form factor $F_{D\pi}(s)$
\begin{eqnarray}
F_{D\pi}(s)=\frac{ \sqrt s f_{D^*} g_{D^*D\pi}}{\mathcal{D}_{\rm BW}(s)}\;.
\label{eqn-def-FDpi}
\end{eqnarray} 
The $f_{D^*}$ above is the decay constant for $D^*$, one can find its different values in Refs~\cite{ijmpa30-1550116,
nppp270-143,epjc75-427,prd96-034524,1810-00296}. We adopt $f_{D^*}=(250\pm11)$ MeV~\cite{ijmpa30-1550116,
nppp270-143} in the numerical calculations. The energy dependent relativistic Breit-Wigner denominator
$\mathcal{D}_{\rm BW}(s)$ equals to $m^2_{D^*}-s-i m_{D^*}\Gamma(s)$, with $m_{D^*}$ the pole mass for $D^*$, 
and the mass dependent decay width defined as
\begin{eqnarray} 
\Gamma(s)=\Gamma_0\left( \frac{q}{q_0}\right)^3\left( \frac{m_{D^*}}{\sqrt s}\right)X^2(qr_{\rm BW}),
\label{eqn-def-gammas}
\end{eqnarray}    
where the barrier radius $r_{\rm BW}=4.0$ GeV$^{-1}$ as it in Refs.~\cite{prd91-092002,prd92-012012,prd94-072001},
the Blatt-Weisskopf barrier factor~\cite{BW-factor1952} is
\begin{eqnarray} 
X(qr_{\rm BW})=\sqrt{\frac{1+(q_0r_{\rm BW})^2}{1+(qr_{\rm BW})^2} },
\label{def-BW-barrier}
\end{eqnarray}
and $q=\frac12\sqrt{\left[s-(m_D+m_\pi)^2\right]\left[s-(m_D-m_\pi)^2\right]/s}$ is the magnitude of the momentum for the 
daughter state $D$ or $\pi$ in the rest frame of the $D^*$, $q_0$ is the value of $q$ at $s=m^2_{D^*}$. For the virtual
contributions from the state $D^{*0}$, the $m_{D^*}$ in the $q_0$ shall be replaced with the $m^\text{eff}_0$, which has its
formula in Refs.~\cite{prd91-092002,prd94-072001,prd90-072003}.

The coupling constant $g_{D^*D\pi}$ in Eq.~(\ref{eqn-def-FDpi}) could be related to the decay width $\Gamma_0$ in 
Eq.~(\ref{eqn-def-gammas}) for $D^*$.
For the total decay width of the $D^{*+}$, which is the sum of the partial widths of the decays
$D^{*+}\to D^0\pi^+, D^{*+}\to D^+\pi^0$ and $D^{*+}\to D^+\gamma$, was firstly measured by CLEO Collaboration 
with $\Gamma(D^{*+})=96\pm4\text{(stat.)}\pm22\text{(syst.)}$ keV~\cite{prd65-032003}. A more precise measurement performed 
by BaBar Collaboration presented $\Gamma(D^{*+})=83.3\pm1.2\text{(stat.)}\pm1.4\text{(syst.)}$ keV and 
$g_{D^*D\pi}=16.92\pm0.13\pm0.14$~\cite{prd88-052003,prl111-111801} with the isospin relation 
$g_{D^*D\pi}=g_{D^*D^0\pi^+}=-\sqrt 2 g_{D^*D^+\pi^0}$. For the state $D^{*0}$, there is no accurate experimental result for 
its decay width. 
In the measurement of three-body decays including virtual $D^{*0}$ contributions, the width was fixed to $0.1$ MeV by BaBar~\cite{prd79-112004}, the experimental upper limit of $2.1$ MeV was adopted by LHCb~\cite{prd91-092002,prd94-072001}, 
while the decay width for $D^{*0}$ in the work~\cite{prd69-112002} from Belle Collaboration was calculated from the width 
of the $D^{*+}$ assuming isospin invariance and HQET.

The Lorentz invariant amplitude ${\mathcal A}$ for the quasi-two-body $B\to D^* h\to D\pi h$ decay processes in
the PQCD approach, according to Fig.~\ref{fig-feyndiag}, is given by~\cite{plb561-258,prd70-054006}
\begin{eqnarray}
{\mathcal A}=\phi_B\otimes H\otimes\phi_{h}\otimes \phi_{D\pi}\;,
\end{eqnarray}
where the symbol $\otimes$ means convolutions in parton momenta, the hard kernel $H$ contains one hard gluon exchange 
%at leading order 
as shown in Fig.~\ref{fig-feyndiag} and the $B$ meson ($h$, $D\pi$ pair) distribution amplitude 
$\phi_B$ ($\phi_{h}$, $\phi_{D\pi}$) absorbs the nonperturbative dynamics in decay processes. The differential branching 
fractions (${\mathcal B}$) for the $B\to D^* h\to D\pi h$ decays are~\cite{PDG-2018,prd74-114009,prd79-094005}
\begin{eqnarray}
\frac{d{\mathcal B}}{d\eta}=\frac{\tau_B q^3_h q^3}{48\pi^3m^5_B}\overline{|{\mathcal A}|^2}\;,
\label{eqn-diff-bra}
\end{eqnarray}
where $\tau_B$ being the $B$ meson mean lifetime. The magnitudes of $h$ meson momentum $q_h$, in the rest frame of 
the $D^*$, is written as
\begin{eqnarray}
q_h=\frac{1}{2}\sqrt{\big[\left(m^2_{B}-m_{h}^2\right)^2 -2\left(m^2_{B}+m_{h}^2\right)s+s^2\big]/s}\;.
\end{eqnarray}
The $m_h$ is the mass of the bachelor meson pion or kaon.
The decay amplitudes for $B\to D^* h\to D\pi h$ are collected in the Appendix.

\section{Results} %XXXXXXXXXXXX-----<<<Results>>>-----

%we adopt the QCD scale $\Lambda_{QCD}=0.25$~GeV.  
In the numerical calculation, we adopt the decay constant $f_B=0.19$ GeV~\cite{prd98-074512}, the mean lifetimes 
$\tau_{B^0}=(1.520\pm0.004)\times 10^{-12}$~s and $\tau_{B^\pm}=(1.638\pm0.004)\times 10^{-12}$~s~\cite{PDG-2018} 
for the $B$ meson. The masses of the neutral and charged $B$, $D$, $\pi$ and $K$ mesons, the pole masses of the neutral 
and charged $D^*$ and the Wolfenstein parameters $\lambda$ and $A$ are presented in Table~\ref{tab1}.
\vspace{-0.2cm}
\begin{table}[thb]
\begin{center}
\caption{Masses (in units of GeV) and Wolfenstein parameters~\cite{PDG-2018}.}
\label{tab1}
\begin{tabular}{l} \hline\hline
\;\;$m_{B^{0}}=5.280 \quad m_{B^{\pm}}=5.279 \quad m_{D^{0}}=1.865 \quad\;  m_{D^{\pm}}=1.870 \quad m_{\pi^0}=0.135$\\
\;\;$m_{\pi^\pm}=0.140 \quad m_{K^0}=0.498 \quad  m_{K^\pm}=0.494 \quad m_{D^{*0}}=2.007 \;\;\; m_{D^{*\pm}}=2.010$
\vspace{0.05cm} \\   \hspace{1.9cm}
%\quad \quad \quad\quad \quad \quad 
$\lambda=0.22453\pm 0.00044  \quad \quad \quad \quad  A=0.836\pm0.015 
% \;\;\;  \bar{\rho} = 0.122^{+0.018}_{-0.017}  \;\;\; %\bar{\eta}= 0.355^{+0.012}_{-0.011}
$\\
\hline\hline
\end{tabular}
\end{center}
\vspace{-0.2cm}
\end{table}

Utilizing the the differential branching fraction the Eq.~(\ref{eqn-diff-bra}) and the decay amplitudes collected in Appendix A, 
we obtain the branching fractions for the virtual contributions (${\mathcal B}_v$) in Table~\ref{tab2} of the concerned 
quasi-two-body decay processes $B\to D^* h\to D\pi h$. 
The invariant mass of the $D\pi$ system has been cut at $2.1$ GeV for the results in Table~\ref{tab2} 
by following the step of Ref.~\cite{prd92-032002}, and the decay width $\Gamma_{D^{*0}}=2.1$ MeV which 
has been adopted by LHCb Collaboration in Refs.~\cite{prd91-092002,prd94-072001} is employed for the two $B^-$ 
decay modes. The largest error for the branching fractions in Table~\ref{tab2} comes from the $B$ meson 
shape parameter uncertainty $\omega_{B}=0.40\pm0.04$ GeV, the error induced by the decay constant $f_{D^*}=(250\pm11)$ 
MeV~\cite{ijmpa30-1550116,nppp270-143} takes the second place, the uncertainty of the Wolfenstein parameter $A$ in 
Table~\ref{tab1} contributes the fourth one, while the third error and the last one originated from the $D^*$ Gegenbauer 
moment $a_{D\pi}=0.50\pm0.10$ and shape parameter $\omega_{D\pi}=0.10\pm0.02$~\cite{prd78-014018,prd90-094018}, 
respectively. There are other errors, which come from the uncertainties of the parameters in the distribution amplitudes for bachelor 
pion(kaon)~\cite{prd86-114025}, the Wolfenstein parameters $\lambda$~\cite{PDG-2018}, etc. are small and have been neglected.
One has the integrated branching ratios for the two-body decays $\bar B^0\to D^{*+}\pi^-$ and %%(\to D^0\pi^+)
$\bar B^0\to D^{*+}K^-$ as  %%(\to D^0\pi^+)  (\to D^0\pi^+)   (\to D^0\pi^+)
\begin{eqnarray}  
\mathcal{B}\left(\bar B^0\to D^{*+}\pi^-\right)&=&\left(2.54^{+1.10}_{-0.76}(\omega_B) ^{+0.23}_{-0.22}(f_{D^*})
^{+0.20}_{-0.22}(a_{D\pi})\pm0.09(A)^{+0.06}_{-0.04}(\omega_{D\pi})\right)\times 10^{-3}\;,
\label{Res-D+pi-}\\ 
\mathcal{B}\left(\bar B^0\to D^{*+}K^-\right)&=&\left(2.05^{+0.81}_{-0.61}(\omega_B) \pm0.18(f_{D^*})
^{+0.15}_{-0.16}(a_{D\pi})\pm0.07(A)\pm0.04(\omega_{D\pi})\right)\times 10^{-4}\;,
\label{Res-D+K-} 
\end{eqnarray}
from the corresponding quasi-two-body decays
by integrating the whole physical region of the $D^0\pi^+$ invariant mass and considering the data
$\mathcal{B}( D^{*+}\to D^0\pi^+)=67.7\%$~\cite{PDG-2018}. 
The two results above predicted by PQCD agree well 
with the branching fractions $(2.74\pm0.13)\times10^{-3}$ and $(2.12\pm0.15)\times10^{-4}$ for the two-body decays 
$\bar B^0\to D^{*+}\pi^-$ and $\bar B^0\to D^{*+}K^-$ in the {\it Review of Particle Physics}~\cite{PDG-2018}, respectively.

%%%%%%%%%%%%%-Table 2-%%PQCD predictions
\begin{table}[thb]
\begin{center}
\caption{The PQCD predictions of the virtual contributions from $D^*$ state in the $D\pi$ invariant mass region $\sqrt s>2.1$ 
               GeV for the $B\to D^* h\to D\pi h$ decays.}
\label{tab2}   
\begin{tabular}{l c l } \hline
    ~~~Mode       &    ~Unit~      &   \;\;\;\;${\mathcal B}_v$ \\  \hline             
  $\bar B^0\;\to D^{*+}\pi^-\;\to D^0\pi^+\pi^-$\;     &$~~~(10^{-4})~~~$
      &\;  $0.87^{+0.43}_{-0.27}(\omega_B) ^{+0.08}_{-0.07}(f_{D^*})^{+0.08}_{-0.06}(a_{D\pi})
     \pm0.03(A)^{+0.02}_{-0.01}(\omega_{D\pi})$\;  \\ 
  $\bar B^0\;\to D^{*+}K^-\to D^0\pi^+K^-$\;     &$~~~(10^{-5})~~~$
      &\; $0.72^{+0.35}_{-0.22}(\omega_B)^{+0.07}_{-0.06}(f_{D^*})\pm0.06(a_{D\pi}) 
     \pm0.03(A)\pm0.02(\omega_{D\pi})$\;  \\ 
  $B^-\to D^{*0}\pi^-\;\;\to D^+\pi^-\pi^-$\;     &$~~~(10^{-4})~~~$
      &\; $1.91^{+0.86}_{-0.59}(\omega_B) ^{+0.17}_{-0.16}(f_{D^*})^{+0.12}_{-0.10}(a_{D\pi}) 
      \pm0.07(A)^{+0.04}_{-0.05}(\omega_{D\pi})$\;  \\    
  $B^-\to D^{*0}K^-\;\to D^+\pi^-K^-$\;     &$~~~(10^{-5})~~~$
      &\; $1.48^{+0.65}_{-0.46}(\omega_B)\pm0.13(f_{D^*})^{+0.09}_{-0.08}(a_{D\pi}) 
     \pm0.05(A)^{+0.02}_{-0.03}(\omega_{D\pi})$\;  \\ 
\hline
\end{tabular} 
\end{center}
\vspace{-0.2cm}
\end{table}
%%%%%%%%%%%%%-Table 2-%%PQCD predictions

For the denominator ${\mathcal{D}_{\rm BW}(s)}$ in the Eq.~(\ref{eqn-def-FDpi}),  we have $|m^2_{D^*}-s|\gg |m_{D^*}\Gamma(s)|$ 
when $\sqrt{s}>2.1$ GeV even if the $D^{*0}$ decay width is $2.1$ MeV. As a result, the variation of the  $r_{\rm BW}$ from 
$4.0$ GeV$^{-1}$ to $1.6$ GeV$^{-1}$~\cite{prd92-032002,prd69-112002} in Eq.~(\ref{eqn-def-gammas}) makes the virtual 
contributions for the $B\to D^* h\to D\pi h$ decays in Table~\ref{tab2} essentially unchanged. The same situation will happen 
again because of the same reason when one replaces Blatt-Weisskopf barrier factor, the Eq.~(\ref{def-BW-barrier}), 
with the exponential form factor (EFF) $F(z)={\rm exp}(-(z-z^\prime))$ for the denominator ${\mathcal{D}_{\rm BW}(s)}$, 
where $z$ and $z^\prime$ have their expressions in Ref.~\cite{prd79-112004}. 
The EFF $R(m^2(D\pi))=e^{-(\beta_1+i\beta_2)m^2(D\pi)}$, with the free parameters $\beta_1$ and $\beta_2$, has been used in 
the experimental Dalitz plot analyses~\cite{prd92-032002} to describe the contributions from the off-shell $D^*(2010)^-$ and 
the general $\bar D^0\pi^-$ P-wave. We don't tend to employ an EFF to replace the time-like form factor of Eq.~(6) because 
the EFF will bring us an unknown parameter and reduce the ability of theoretical prediction.
As a test of the effect of $m^\text{eff}_0$ instead of $m_{D^{*0}}$ for the virtual contributions 
in the decays involving $D^{*0}$ in this work, 
we employ the value $(m_0^{\rm eff}+\Delta m)$ to replace the $m^\text{eff}_0$ for $q_0$ in the Eq.~(\ref{def-BW-barrier}).
When $\Delta m$ is $\pm0.5$ GeV or even $\pm1.0$ GeV, the results for $B^-\to D^{*0}\pi^-\to D^+\pi^-\pi^-$ and
$B^-\to D^{*0}K^-\to D^+\pi^-K^-$ are almost the same as they in Table~\ref{tab2}, the variations are found
less than 0.1\% for the corresponding values.

%%%%%%%%%%%%%%%%%%%%%%%%%%%%%%%%%%%%%%%%%%%%%%%%%%
\begin{figure}[H] %[tbp] quasi-two-body 
%\vspace{-1cm}
\centerline{\epsfxsize=7.0cm \epsffile{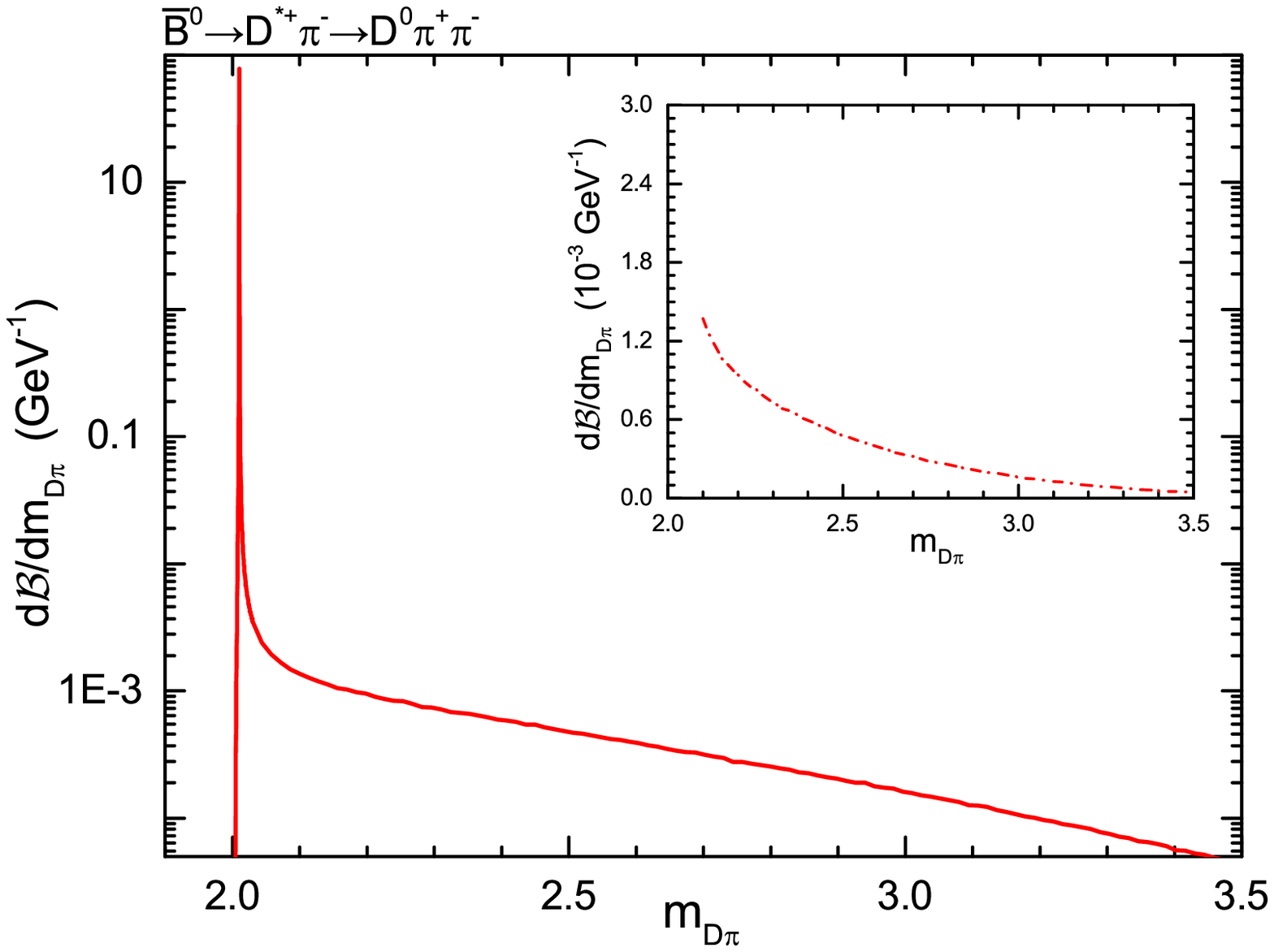}
                  \epsfxsize=7.0cm \epsffile{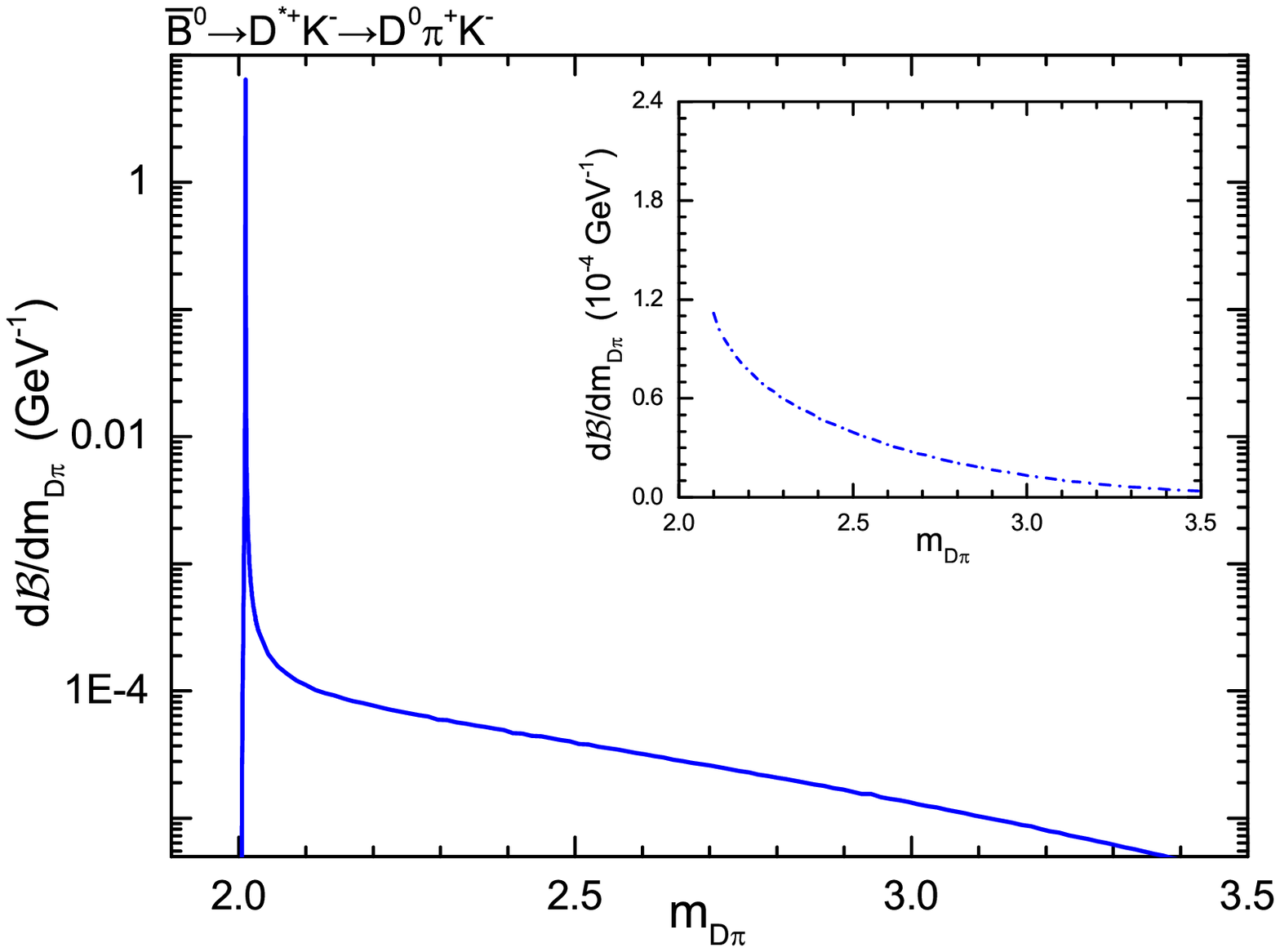}}
\vspace{-0.5cm}
\caption{The differential branching fractions for the decays $\bar B^0\to D^{*+}\pi^-\to D^0\pi^+\pi^-$ 
              (left) and $\bar B^0\to D^{*+}K^-\to D^0\pi^+K^-$ (right). The two small diagrams are for the corresponding 
              virtual contributions in the $m_{D\pi}$ region ($2.1\sim3.5$) GeV.}
\label{fig-Dplus}
\vspace{-0.2cm}
\end{figure}
\begin{figure}[H] %[tbp]dep-D2400
%\vspace{-1cm}
\centerline{%\epsfxsize=6.0cm \epsffile{dep-D2400.eps}
                  \epsfxsize=7.0cm \epsffile{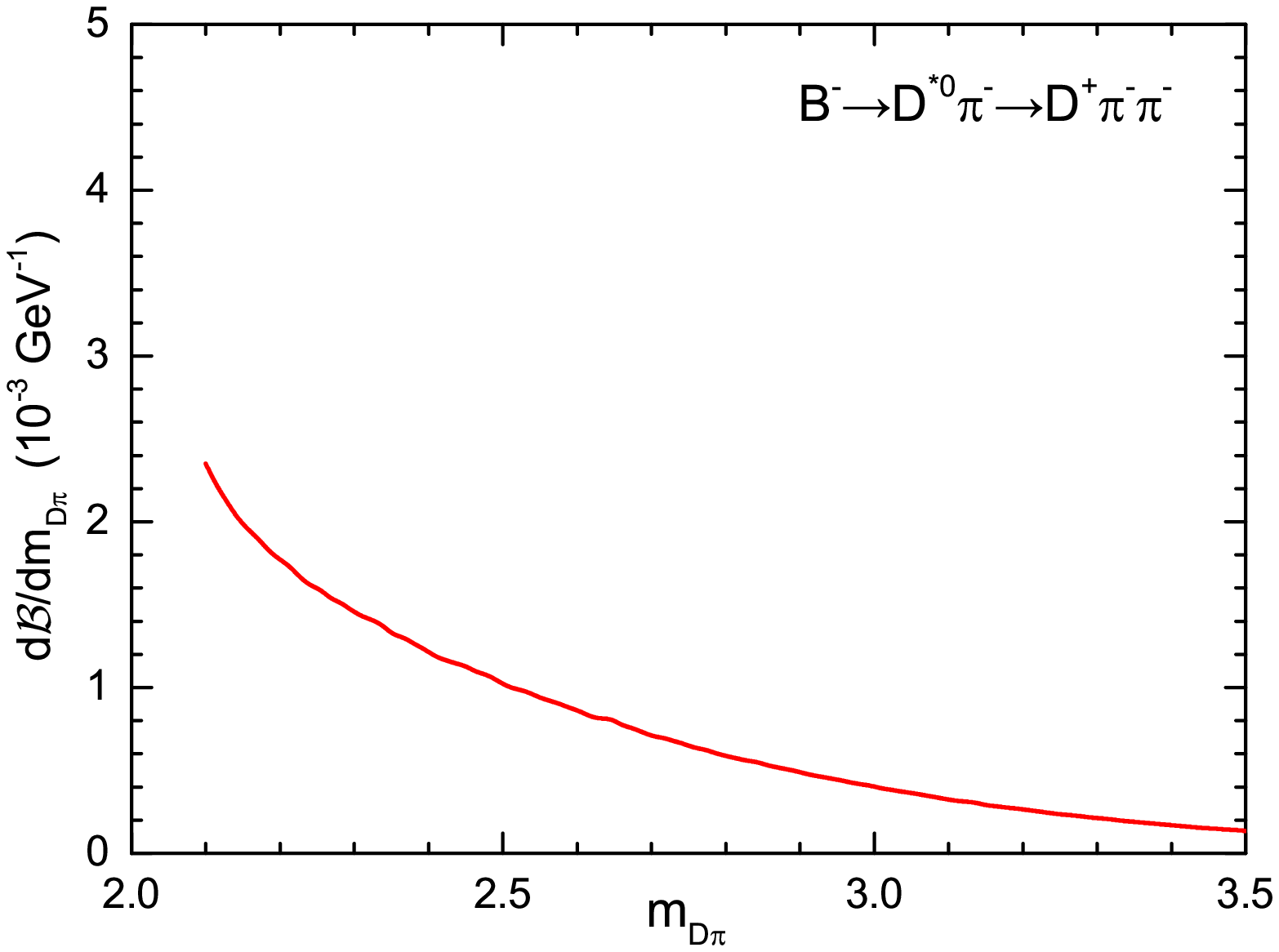}
                  \epsfxsize=7.0cm \epsffile{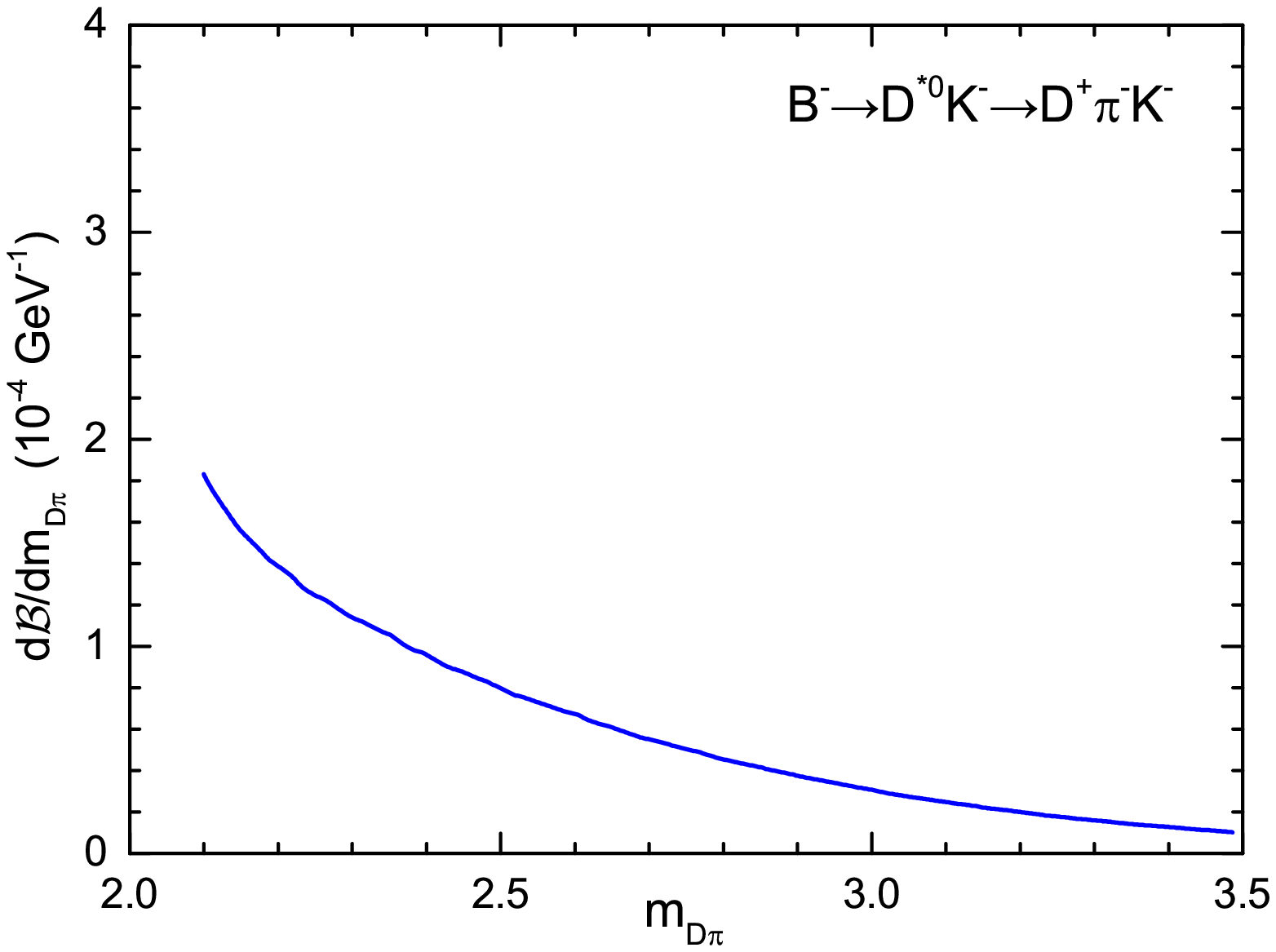}}
\vspace{-0.5cm}
\caption{The differential branching fractions for the processes $B^-\to D^{*0}\pi^-\to D^+\pi^-\pi^-$ (left) 
              and $B^-\to D^{*0}K^-\to D^+\pi^-K^-$ (right) in the $m_{D\pi}$ region ($2.1\sim3.5$) GeV.}
\label{fig-Dzero}
\vspace{-0.2cm}
\end{figure}

%%%%%%%%%%%%%%%%%%%%%%%%%%%%%%%%%%%%%%%%%%%%%%%%%%
The distributions of the branching ratios for the quasi-two-body decays $\bar B^0\to D^{*+}\pi^-\to D^0\pi^+\pi^-$ and 
$\bar B^0\to D^{*+}K^-\to D^0\pi^+K^-$ in the $D\pi$ pair invariant mass $m_{D\pi}$ (equals to $\sqrt s$) are shown 
in Fig.~\ref{fig-Dplus}.
These diagrams reveal that the main portion of the values of Eq.~(\ref{Res-D+pi-}) and Eq.~(\ref{Res-D+K-}) 
concentrate in a very small region of the $D^0\pi^+$ invariant mass. 
Take $\bar B^0\to D^{*+}K^-\to D^0\pi^+K^-$ as an example, 
we have $\{92.7\%, 92.9\%, 93.6\%\}$ of its quasi-two-body branching ratio in the $m_{D^0\pi^+}$ region 
$\left(m_{D^{*+}}-\delta_m, m_{D^{*+}}+\delta_m\right)$ when $\delta_m=\{2.5, 3.5, 5.0\}$ MeV.
The distinct feature of the diagrams with two very sharp peaks located at the mass of $D^{*+}$ in Fig.~\ref{fig-Dplus} is 
%for  the decays $\bar B^0\to D^{*+}\pi^-\to D^0\pi^+\pi^-$ and $\bar B^0\to D^{*+}K^-\to D^0\pi^+K^-$ originates from
dominated by the tiny decay width $\Gamma(D^{*+})=83.3$ keV~\cite{prd88-052003,prl111-111801} for $D^{*+}$ and the 
$D^0\pi^+$ threshold which is so close to $m_{D^{*+}}$.
%\pm1.2\text{(stat.)}\pm1.4\text{(syst.)}
These two points result in dramatic difference of the curves when comparing with the differential branching fractions 
predicted in Ref.~\cite{1809-02943} for the $B\to D^*_0(2400) h\to D\pi h$ decays with the broad resonant state.
%The narrow peaks in Fig.~\ref{fig-Dplus} cause the proportion of the branching fractions for the processes including 
%the $D^*$ as the intermediate state important in the region far from the pole mass of $D^{*+}$ which could be also concluded from 
The differential branching fractions for the $B^-\to D^{*0}\pi^-\to D^+\pi^-\pi^-$ and $B^-\to D^{*0}K^-\to D^+\pi^-K^-$ 
are shown in Fig.~\ref{fig-Dzero}. The $d\mathcal{B}/dm_{D\pi}$ values at the point $m_{D\pi}=3.5$ GeV are about $5\%$ of the  
values at $m_{D\pi}=2.1$ GeV for both the decays $B^-\to D^{*0}\pi^-\to D^+\pi^-\pi^-$ and $B^-\to D^{*0}K^-\to D^+\pi^-K^-$.
%While for the decay $B^-\to D^{*}_0(2400)^0K^-\to D^+\pi^-K^-$, it's the point $m_{D\pi}=2.85$ GeV which got  $\frac{1}{13\%}$ 
%of $d\mathcal{B}/dm_{D\pi}$ the value at $m_{D\pi}=3.5$ GeV for the $D\pi$ pair decay from $D^{*}_0(2400)^0$ in 
%Ref.~\cite{1809-02943}.
%%%%%%%%%%%%%%%%%%%%%%%%%%%%%%%%%%%%%%%%%%%%%%%%%%
%%%%%%%%%%%%%%%%%%%%%%%%%%%%%%%%%%%%%%%%%%%%%%%%%%
%%%%%%%%%%%%%-Table 3-%%comparision
\begin{table}[thb]
\begin{center}  %%$\sqrt s>2.1$ GeV for the $B\to D^* h\to D\pi h$ decays. errors are added in quadrature.
\caption{The comparison of the predicted virtual contributions with the experimental measurements.}
\label{tab3}   
\begin{tabular}{l l l l} \hline
    ~~~Mode          & ~ \;\;\;\;${\mathcal B}_v$  &    \;\;\;\;Data~ & Ref.\;  \\  \hline             
  $\bar B^0\;\to D^{*+}\pi^-\;\to D^0\pi^+\pi^-$   
      &\;  $0.87^{+0.45}_{-0.29}\times 10^{-4} $\;\;    &\;$(0.88\pm0.13)\times 10^{-4}$  &\;\cite{prd76-012006} \\ 
   ~    &~                                                                  &\;$\sim0.78\times 10^{-4}$  &\;\cite{prd92-032002} \\ 
  $\bar B^0\;\to D^{*+}K^-\to D^0\pi^+K^-$   
      &\; $0.72^{+0.36}_{-0.24}\times 10^{-5}$\;\;   &\;$(0.81\pm0.15\pm0.20\pm0.27\pm0.09)\times 10^{-5}$\;  
                      																	&\;\cite{prd92-012012}\\ 
  $B^-\to D^{*0}\pi^-\;\;\to D^+\pi^-\pi^-$    
      &\; $1.91^{+0.89}_{-0.63}\times 10^{-4}$\;\;      &\;$(2.23\pm0.32)\times 10^{-4}$  &\;\cite{prd69-112002}\\    
  ~    & ~                                                                  &\;$\sim1.09\times 10^{-4}$  &\;\cite{prd79-112004}\\    
  ~    & ~                                                                  &\;$(1.09\pm0.07\pm0.07\pm0.24\pm0.07)\times 10^{-4}$\;
  																		  &\;\cite{prd94-072001}\\    
  $B^-\to D^{*0}K^-\;\to D^+\pi^-K^-$    
      &\; $1.48^{+0.67}_{-0.49}\times 10^{-5}$\;\;      &\;$(5.6\pm1.7\pm1.0\pm1.1\pm0.4)\times 10^{-6}$  &\;\cite{prd91-092002}\\ 
\hline
\end{tabular} 
\end{center}
\vspace{-0.2cm}
\end{table}
%%%%%%%%%%%%%-Table 2-%%PQCD predictions
The comparison of the predicted virtual contributions with the experimental measurements are presented 
in the Table~\ref{tab3}, the theoretical errors are added in quadrature.
For the $\bar B^0\to D^{*+}\pi^-\to D^0\pi^+\pi^-$ decay, the branching fraction of the two-body subprocess 
$\bar B^0\to D^{*+}\pi^-$ was found to be $(2.22\pm0.04\pm0.19)\times 10^{-3}$, which is about $87\%$ of the 
Eq.~(\ref{Res-D+pi-}) and $81\%$ of the corresponding data in~\cite{PDG-2018} for the central value,  
in the $m_{D^0\pi^+}$ region within $3$ MeV of the nominal $D^{*+}-D^0$ mass difference in Ref.~\cite{prd76-012006} 
by Belle Collaboration, and the relevant quasi-two-body virtual contribution is $(0.88\pm0.13)\times 10^{-4}$.
In the Ref.~\cite{prd92-032002} for the same decay process, with $m_{D^0\pi^+}>2.1$ GeV, 
the $D^0\pi^+$ $P$-wave contribution is $(9.21\pm0.56\pm0.24\pm1.73)\%$ (isobar model) and 
$(9.22\pm0.58\pm0.67\pm0.75)\%$ (K-matrix model) of the total branching fraction 
$(8.46\pm0.14\pm0.29\pm0.40)\times 10^{-4}$, that is about $0.78\times 10^{-4}$ as shown in Table~\ref{tab3}, 
which is close to the PQCD prediction.
For the decay $\bar B^0\to D^{*+}K^-\to D^0\pi^+K^-$, in Ref.~\cite{prd92-012012}, 
within $2.5$ MeV of the $D^{*+}-D^0$ mass difference to remove background containing $D^{*+}\to D^0\pi^+$, 
the virtual contribution was treated as part of the background with $D\pi$ P-wave nonresonant contributions
as a result of $(0.81\pm0.15\pm0.20\pm0.27\pm0.09)\times 10^{-5}$ 
which is slightly larger than the corresponding result in Table~\ref{tab3}. 
For the $B^-\to D^{*0}\pi^- \to D^+\pi^-\pi^-$ decay,  the Belle Collaboration provided $(2.23\pm0.32)\times 10^{-4}$ for the 
branching fraction of the virtual contribution in~\cite{prd69-112002} with the parameterization 
$F(q)={\rm exp}\left(-r_{\rm BW}(q-q_0)\right)$ for $D^{*0}\to D^+$ form factor. The same form factor was adopted for the same
decay process in~\cite{prd79-112004} by BaBar Collaboration, while $(10.1\pm1.4)\%$ of the total branching fraction 
$(1.08\pm0.03)\times 10^{-3}$, about $1.09\times 10^{-4}$, 
was obtained for the same virtual contribution, which is about half of the corresponding 
result in Table~\ref{tab3}. In Ref.~\cite{prd94-072001}, LHCb presented the experimental result 
$(1.09\pm0.07\pm0.07\pm0.24\pm0.07)\times 10^{-4}$ for the same virtual contribution. 
As for the $B^-\to D^{*0}K^- \to D^+\pi^-K^-$ decay, LHCb presented the experimental result 
$(5.6\pm1.7\pm1.0\pm1.1\pm0.4)\times 10^{-6}$ in Ref.~\cite{prd91-092002}
for the virtual contribution which is only about $1/3$ of the corresponding PQCD prediction. 

\begin{figure}[H] %[tbp]
%\vspace{-1cm}
\centerline{\epsfxsize=7.0cm \epsffile{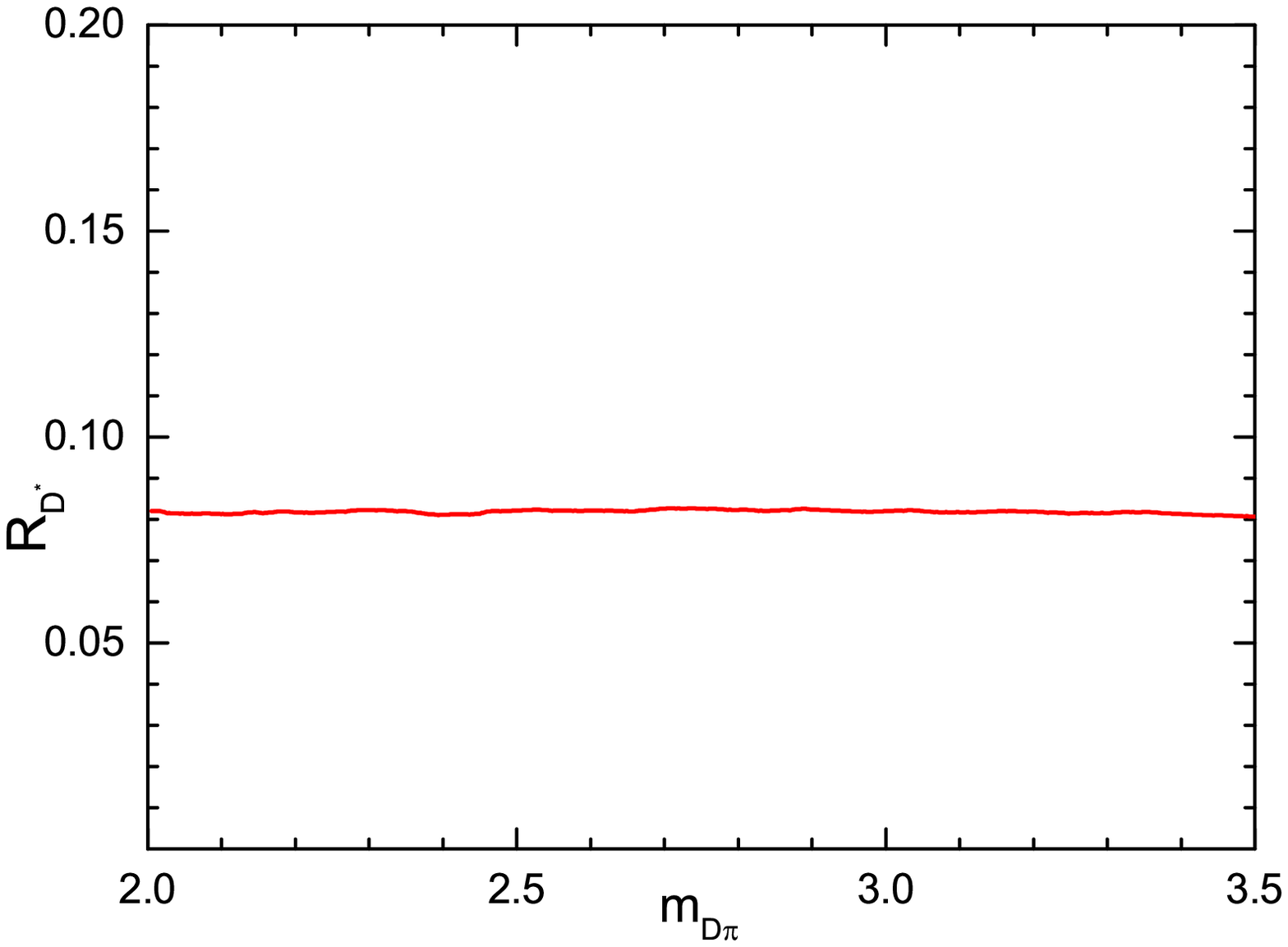}
                  \epsfxsize=7.0cm \epsffile{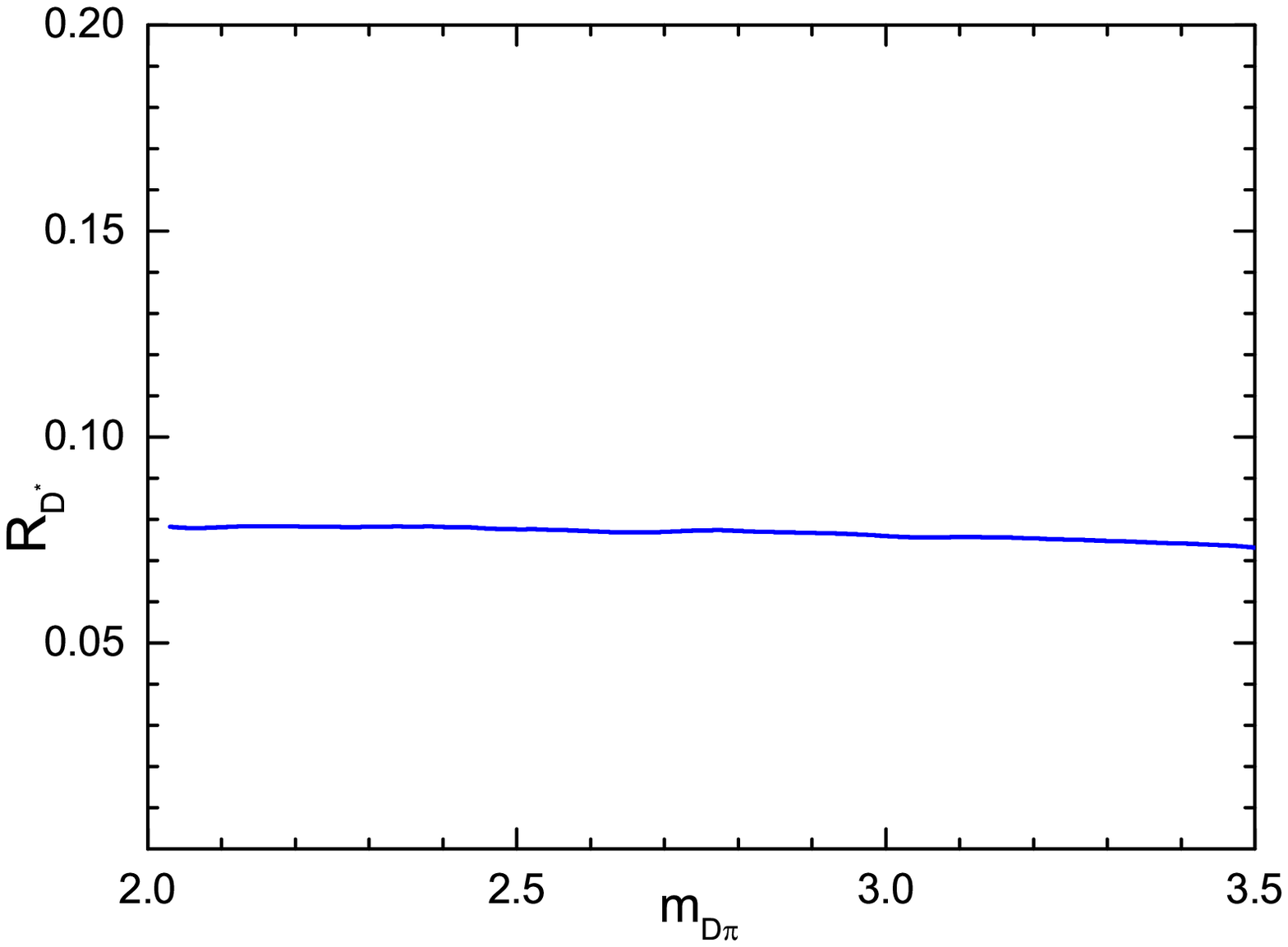}}
\vspace{-0.5cm}
\caption{The energy dependent ratios $R_{D^{*+}}$ (left) and $R_{D^{*0}}$ in the region ($2.03\sim3.50$) GeV (right) 
               predicted by PQCD.}
\label{fig-RD}
\vspace{-0.2cm}
\end{figure}

For the quasi-two-body processes $\bar B^0\to D^{*+}K^-\to D^0\pi^+K^-$ and $\bar B^0\to D^{*+}\pi^-\to D^0\pi^+\pi^-$,
we have an identical step $D^{*+}\to D^0\pi^+$, the difference of these two decay modes originated from the bachelor particles 
pion and kaon. Assuming factorization and flavor-$SU(3)$ symmetry, one has the ratio $R_{D^{*+}}$ for the branching fractions 
of these two processes as
\begin{eqnarray}
R_{D^{*+}}=\frac{{\mathcal{B}(\bar B^0\to D^{*+}K^-\to D^0\pi^+K^-)}}{{\mathcal{B}(\bar B^0\to D^{*+}\pi^-\to D^0\pi^+\pi^-)}}
\approx \left|\frac{V_{us}}{V_{ud}}\right|^2\cdot \frac{f^2_K}{f^2_\pi}.
\label{def-RDplus}
\end{eqnarray}
With the result
\begin{eqnarray}
\left|\frac{V_{us}}{V_{ud}}\right|\frac{f_{K^+}}{f_{\pi^+}}=0.276
\end{eqnarray}
in {\it Review of Particle Physics}~\cite{PDG-2018}, one has $R_{D^{*+}}\approx0.076$.
The PQCD predicted branching ratios %%, the Eq.~(\ref{Res-D+pi-}) and Eq.~(\ref{Res-D+K-}), 
%%for the $\bar B^0\to D^{*+}K^-\to D^0\pi^+K^-$ and $\bar B^0\to D^{*+}\pi^-\to D^0\pi^+\pi^-$ decays 
provide 
\begin{eqnarray} 
R_{D^{*+}}=0.081^{+0.000}_{-0.002}(\omega_B)^{+0.001}_{-0.000}(a_{D\pi}).
\label{res-RD+}
\end{eqnarray}
It's clear that the breaking effects of the flavor-$SU(3)$ symmetry is quite small for $R_{D^{*+}}$.~The small errors induced by 
the uncertainties of $\omega_B$ and $a_{D\pi}$ for $R_{D^{*+}}$ are caused by the cancellation, 
which means the increase or decrease for the values of these parameters will result in nearly identical change of the 
weight at the same direction for the branching ratios of these two decays.  And the errors of 
$R_{D^{*+}}$ come from the $f_{D^*}$, Wolfenstein parameter $A$  and $\omega_{D\pi}$ are zeros for the same reason. 
The result of Eq.~(\ref{res-RD+}) is consistent with the data $(7.76\pm0.34\pm0.29)\%$ presented by 
BaBar~\cite{prl96-011803} and $(0.074 \pm 0.015 \pm 0.006)$ announced by Belle~\cite{prl87-111801}.
The energy dependent $R_{D^{*+}}$ is shown as the left diagram in Fig.~\ref{fig-RD}.
A similar ratio $R_{D^{*0}}$, which has the definition as
\begin{eqnarray}
R_{D^{*0}}=\frac{{\mathcal{B}(B^-\to D^{*0}K^-\to D^+\pi^-K^-)}}{{\mathcal{B}(B^-\to D^{*0}\pi^-\to D^+\pi^-\pi^-)}}
\approx \left|\frac{V_{us}}{V_{ud}}\right|^2\cdot \frac{f^2_K}{f^2_\pi}\;, 
\label{def-RD0}
\end{eqnarray}%%%%%%%9999999999%%%%%%
for the quasi-two-body decays $B^-\to D^{*0}K^-\to D^+\pi^-K^-$ and $B^-\to D^{*0}\pi^-\to D^+\pi^-\pi^-$
is shown as the right diagram of Fig.~\ref{fig-RD} in the $m_{D\pi}$ region ($2.03\sim3.50$) GeV.
An interesting conclusion could be made from the $R_{D^{*}}$ lines in Fig.~\ref{fig-RD} is that the flavor-$SU(3)$ 
symmetry will be maintained at any physical point of the invariant mass $m_{D\pi}$ for the concerned quasi-two-body 
$B\to D^* h\to D\pi h$ decays.
The ratio between the branching fractions of the two-body decays $B^-\to D^{*0}K^-$ and $B^-\to D^{*0}\pi^-$ were measured to be
$(7.930 \pm0.110{\rm (stat)} \pm0.560{\rm (syst)})\times 10^{-2}$ by LHCb~\cite{plb777-16} and
$0.0813\pm0.0040{\rm(stat)}^{+0.0042}_{-0.0031}{\rm(syst)}$
at BaBar~\cite{prd71-031102}, which are close to the result
\begin{eqnarray} 
R_{D^{*0}}=0.077^{+0.000}_{-0.001}(\omega_B)^{+0.000}_{-0.001}(\omega_{D\pi}).
\label{res-RD0}
\end{eqnarray}
in the region ($2.1\sim3.5$) GeV deduced from the results in Table~\ref{tab2}.  
%%%%%%%%%%%%%%%%%%%%%%%%%%%%%%%%%%%%%%%%%%%%%%%%%%
%%%%%%%%%%%%%%%%%%%%%%%%%%%%%%%%%%%%%%%%%%%%%%%%%%
The PQCD predictions of the virtual contributions in Table~\ref{tab2} for the $\bar B^0\to D^{*+}\pi^-\to D^0\pi^+\pi^-$ and 
$\bar B^0\to D^{*+}K^-\to D^0\pi^+K^-$ decays are $3.4\%$ and $3.5\%$ in the $D\pi$ invariant mass region $\sqrt s>2.1$ 
GeV from the intermediate state $D^{*+}$, respectively, of the corresponding two-body decay branching ratios, 
the Eq.~(\ref{Res-D+pi-}) and Eq.~(\ref{Res-D+K-}). 
By considering $\mathcal{B}( D^{*+}\to D^0\pi^+)=67.7\%$~\cite{PDG-2018}, we have about $5\%$ 
of the total quasi-two-body branching ratios for the virtual contributions of the two decay processes involving $D^{*+}$.
The virtual contributions in Table~\ref{tab2} for the two $B^-$ decay modes are $3.9\%$ and $3.7\%$ of the
two-body data for $B^-\to D^{*0}\pi^-$ and $B^-\to D^{*0}K^-$ in~\cite{PDG-2018}, respectively.
Because of the threshold of $D^+\pi^-$, we don't have the integrated quasi-two-body branching fractions for the decays 
$B^-\to D^{*0}\pi^-\to D^+\pi^-\pi^-$ and $B^-\to D^{*0}K^-\to D^+\pi^-K^-$.
But we can analyse the quasi-two-body processes $B^-\to D^{*0}\pi^-\to D^0\pi^0\pi^-$ 
and $B^-\to D^{*0}K^-\to D^0\pi^0K^-$ from the $D^0\pi^0$ threshold. 
As an extreme example, if the experimental upper limit of $2.1$ MeV is used for the $D^{*0}$ width,
we have $\mathcal{B}(B^-\to D^{*0}\pi^-)=2.69\times 10^{-4}$ and 
$\mathcal{B}(B^-\to D^{*0}K^-)=2.10\times 10^{-5}$ as the central values for the these two decays
after take into the factor $\mathcal{B}( D^{*0}\to D^0\pi^0)=64.7\%$~\cite{PDG-2018}.
Obviously, the branching fractions for the two-body decays $B^-\to D^{*0}\pi^-$ and 
$B^-\to D^{*0}K^-$ are highly underestimated with $\Gamma_{D^{*0}}=2.1$ MeV.

%%%%%%%%%%%%%%%%%%%%%%%%%%%%%%%%%%%%%%%%%%%%%%%%%%
\begin{figure}[H] %[tbp]
%\vspace{-1cm}
\centerline{\epsfxsize=7.0cm \epsffile{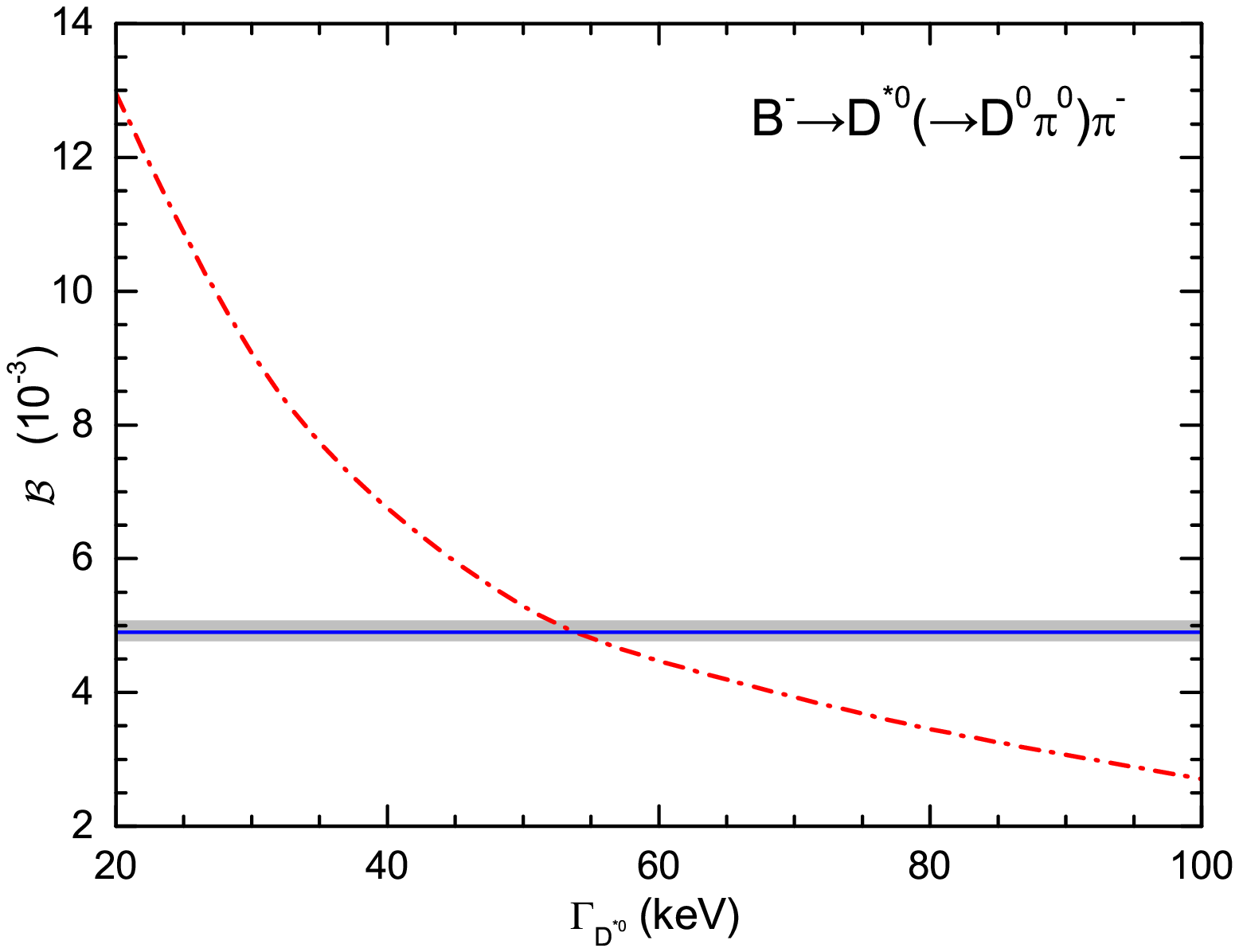}
                  \epsfxsize=7.0cm \epsffile{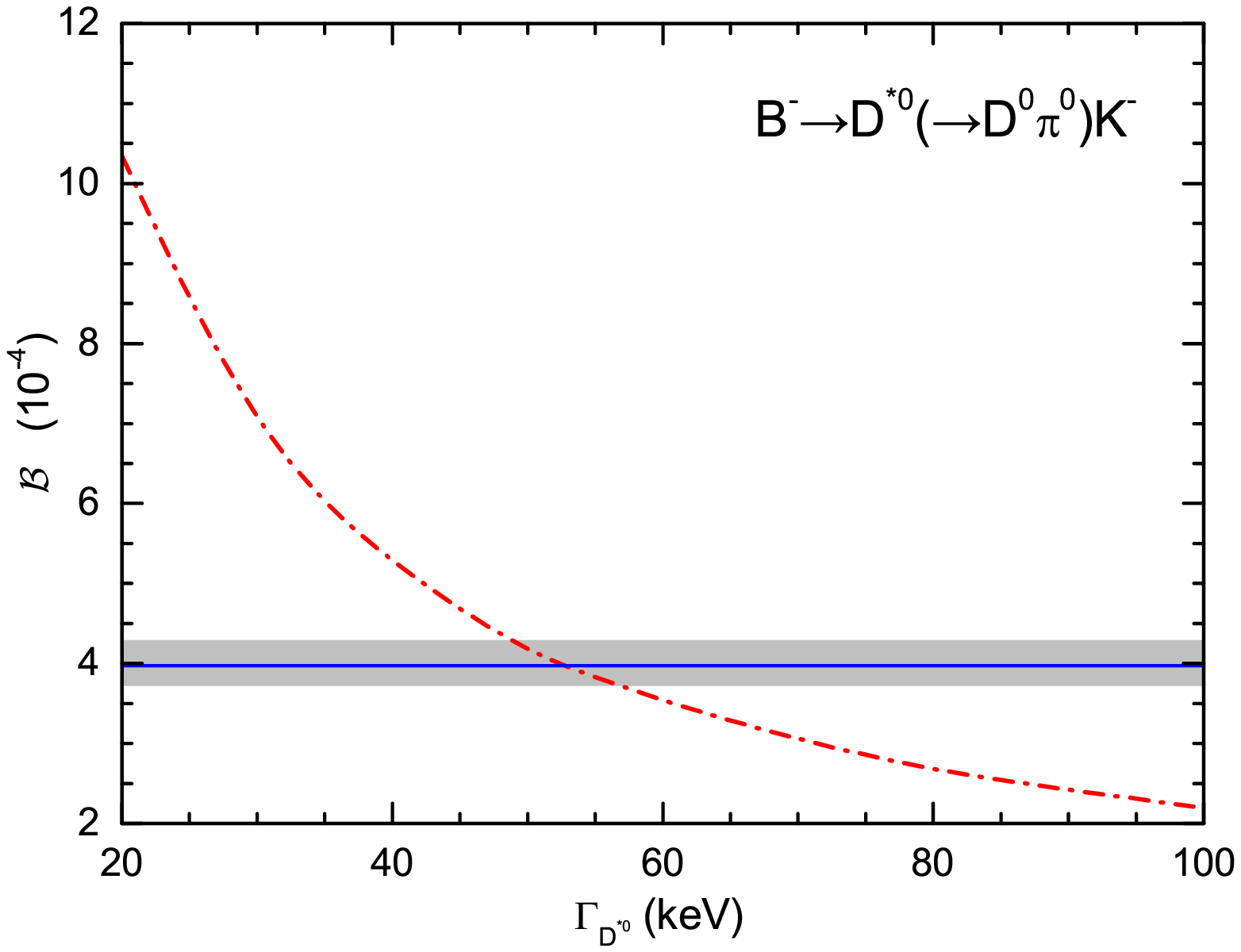}}
\vspace{-0.5cm}
\caption{The $\Gamma_{D^{*0}}$ dependent branching ratios of the two-body decays $B^-\to D^{*0}\pi^-$ (left) and 
              $B^-\to D^{*0}K^-$ (right) with the subprocess $D^{*0}\to D^0\pi^0$. The dash-dot curves are the PQCD predictions, 
              the blue lines and the gray bands are the data with their errors from~\cite{PDG-2018}.}
\label{fig-Gamma}
\vspace{-0.2cm}
\end{figure}
%%%%%%%%%%%%%%%%%%%%%%%%%%%%%%%%%%%%%%%%%%%%%%%%%%

Although there is no direct measurement for the  $\Gamma_{D^{*0}}$, we have theoretical results
$58$ keV~\cite{prd78-014029}, $59.6\pm1.2$ keV~\cite{jhep1404-177}, $55.9\pm1.6$ keV~\cite{prd88-034034} and
$53 \pm5\pm7$ keV~\cite{plb721-94} for it. If we replace the decay width $2.1$ MeV with $\Gamma_{D^{*0}}=53$ 
keV~\cite{plb721-94} and considering the factor $\mathcal{B}( D^{*0}\to D^0\pi^0)=64.7\%$~\cite{PDG-2018}, 
we have %the central values
\begin{eqnarray}
\mathcal{B}(B^-\to D^{*0}\pi^-)&=&\left(5.03^{+2.28}_{-1.44}(\omega_B) ^{+0.45}_{-0.43}(f_{D^*})
^{+0.31}_{-0.23}(a_{D\pi})\pm0.18(A)^{+0.18}_{-0.28}(\omega_{D\pi})\right)\times 10^{-3}\;, \\
\label{eqn-br-D0pi0pi}
\mathcal{B}(B^-\to D^{*0}K^-)&=&\left(3.99^{+1.75}_{-1.21}(\omega_B) ^{+0.36}_{-0.34}(f_{D^*})
^{+0.17}_{-0.24}(a_{D\pi})\pm0.14(A)^{+0.09}_{-0.078}(\omega_{D\pi})\right)\times 10^{-4}\;,
\label{eqn-br-D0pi0K}
\end{eqnarray}
as the two-body branching ratios, which are consistent with the data $\mathcal{B}(B^-\to D^{*0}\pi^-)=(4.90\pm0.17)\times10^{-3}$, 
$\mathcal{B}(B^-\to D^{*0}K^-)=3.97^{+0.31}_{-0.28}\times10^{-4}$~\cite{PDG-2018,prd97-012005,plb777-16}.
The virtual contributions with $\Gamma_{D^{*0}}=53$ keV~\cite{plb721-94} are
\begin{eqnarray}
\mathcal{B}_v(B^-\to D^{*0}\pi^-\to D^0\pi^0\pi^-)&=&\left(0.99^{+0.45}_{-0.30}(\omega_B) \pm0.09(f_{D^*})
^{+0.06}_{-0.05}(a_{D\pi})\pm0.04(A)^{+0.02}_{-0.01}(\omega_{D\pi})\right)\times 10^{-4}\;, \\
\label{eqn-vbr-D0pi0pi}
\mathcal{B}_v(B^-\to D^{*0}K^-\to D^0\pi^0K^-)&=&\left(0.77^{+0.34}_{-0.24}(\omega_B) \pm0.07(f_{D^*})
^{+0.04}_{-0.05}(a_{D\pi})\pm0.03(A)^\pm0.01(\omega_{D\pi})\right)\times 10^{-5}\;,
\label{eqn-vbr-D0pi0K}
\end{eqnarray}
in the region $m_{D^0\pi^0}>2.1$ GeV. The percentage of the virtual contributions are all about $3\%$
of the corresponding quasi-two-body branching ratios.
The $\Gamma_{D^{*0}}$ dependent branching ratios of $B^-\to D^{*0}\pi^-$ and $B^-\to D^{*0}K^-$ with the subprocess
$D^{*0}\to D^0\pi^0$ are shown in Fig.~\ref{fig-Gamma}, the dash-dot curves are the PQCD predictions,  
the blue lines and the gray bands are the data with their errors from~\cite{PDG-2018}. The branching ratios 
%on the $\Gamma_{D^{*0}}$
in Fig.~\ref{fig-Gamma} can be exploited to constrain the $D^{*0}$  decay width which could be read as 
$\Gamma_{D^{*0}}\approx 53$ keV from these two diagrams. The detailed discussion including the impacts 
of different parameter uncertainties about $D^{*0}$ decay width in the three-body hadronic $B$ meson 
decays shall be left for the future study. 
It must be pointed out that, the changes are tiny for the two virtual contributions involving $D^{*0}$ in the Table~\ref{tab2} 
when we adopt $53$ keV for $\Gamma_{D^{*0}}$, the reason is that the $|m_{D^*}\Gamma(s)|$ shall be less than
$1/10^{4}$ of $|m^2_{D^*}-s|$ even if $\Gamma_{D^{*0}}$ equals to $2.1$ MeV when $m_{D^+\pi^-}$ is larger than $2.1$ GeV.
The small branching fractions for the two-body decays $B^-\to D^{*0}\pi^-$ and $B^-\to D^{*0}K^-$ with the subprocess 
$D^{*0}\to D^0\pi^0$ in this work with $\Gamma_{D^{*0}}=2.1$ MeV is caused by the insufficient contributions in the  
$m_{D^0\pi^0}$ region near the $D^{*0}$ pole mass but not by the lower virtual contributions in the region 
$m_{D^0\pi^0}>2.1$ GeV.

\section{CONCLUSION}
In this work, we studied the quasi-two-body decays $B\to D^*h \to D\pi h$ and focused on the virtual contributions
originated from off-shell $D^{\ast}(2007)^0$ and $D^{\ast}(2010)^{\pm}$ in the decays of $B^-\to D^+\pi^-\pi^-$, 
$B^- \to D^+\pi^-K^-$, $\bar B^0\to D^0\pi^+\pi^-$ and $\bar B^0\to D^0\pi^+K^-$
which have been measured by Belle, BaBar and LHCb Collaborations.
For the $\bar B^0 \to D^{*+}\pi^-\to D^0\pi^+\pi^-$ and $\bar B^0\to D^{*+}K^-\to D^0\pi^+K^-$ decays, we found that the 
main portions of their quasi-two-body branching fractions concentrate in a very small region of the $D^0\pi^+$ invariant mass, 
the percentage is larger than $90\%$ for the branching ratios in the realm of $2.5$ MeV around the pole mass of $D^{*+}$.
And the virtual contributions from $D^{*+}$, in the region of $m_{D^0\pi^+}>2.1$ GeV, are about $5\%$ of the integrated values 
for the corresponding quasi-two-body results by considering $\mathcal{B}( D^{*+}\to D^0\pi^+)=67.7\%$.
The virtual contributions in this work for $B^-\to D^{*0}\pi^- \to D^+\pi^-\pi^-$ and 
$B^- \to D^{*0}K^- \to D^+\pi^-K^-$ decay modes were found to be $3.9\%$ and $3.7\%$ of the
two-body data for $B^-\to D^{*0}\pi^-$ and $B^-\to D^{*0}K^-$ in {\it Review of Particle Physics}, respectively.

From the ratios  $R_{D^{*+}}$ and  $R_{D^{*0}}$ defined between the quasi-two-body decays including $D^{*+}$ and $D^{*0}$
as the intermediate states, respectively, we concluded that the flavor-$SU(3)$ symmetry will be maintained with very small 
breaking at any physical value of the invariant mass $m_{D\pi}$ for the concerned $B\to D^* h\to D\pi h$ decays.
We found that the decays $B^-\to D^{*0}\pi^-\to D^0\pi^0\pi^-$ and $B^-\to D^{*0}K^-\to D^0\pi^0K^-$ have strong 
dependence on the $D^{*0}$ decay width for their branching fractions which could be employed as a constraint for 
$\Gamma_{D^{*0}}$.  $2.1$ MeV for $D^{*0}$ decay width will make the branching ratios of the quasi-two-body decays
$B^-\to D^{*0}\pi^-\to D^0\pi^0\pi^-$ and $B^-\to D^{*0}K^-\to D^0\pi^0K^-$ be highly underestimated 
because of the insufficient contributions in the  $m_{D^0\pi^0}$ region near the $D^{*0}$ pole mass. 
With $\Gamma_{D^{*0}}=53$ keV, we predicted
\begin{eqnarray}
\mathcal{B}_v(B^-\to D^{*0}\pi^-\to D^0\pi^0\pi^-)&=&\left(0.99^{+0.45}_{-0.30}(\omega_B) \pm0.09(f_{D^*})
^{+0.06}_{-0.05}(a_{D\pi})\pm0.04(A)^{+0.02}_{-0.01}(\omega_{D\pi})\right)\times 10^{-4}\;, \\
\label{eqn-vbr-D0pi0pi}
\mathcal{B}_v(B^-\to D^{*0}K^-\to D^0\pi^0K^-)&=&\left(0.77^{+0.34}_{-0.24}(\omega_B) \pm0.07(f_{D^*})
^{+0.04}_{-0.05}(a_{D\pi})\pm0.03(A)^\pm0.01(\omega_{D\pi})\right)\times 10^{-5}\;,
\label{eqn-vbr-D0pi0K}
\end{eqnarray}
as the virtual contributions, which are about  $3\%$ of the corresponding quasi-two-body results.

%%%--AAAAA---\begin{eqnarray} \end{eqnarray}
%-----------------------%
\begin{acknowledgments}
We thank Tao Zhong for valuable discussions. 
This work was supported in part by National Science Foundation of China under Grant No. 11547038.
\end{acknowledgments}
%-----------------------%

\appendix

\section{DECAY AMPLITUDES}
The concerned quasi-two-body decay amplitudes are given, in the PQCD approach, by
\begin{eqnarray}
{\mathcal A}\big( \bar B^0\to \pi^-[ D^*(2010)^+\to] D^0\pi^+\big)&=&\frac{G_F}{\sqrt2}V_{cb}V^*_{ud}
\big[\big(\frac{c_1}{3}+c_2\big)F_{TD^*} +c_1M_{TD^*} +\big(c_1+\frac{c_2}{3}\big) F_{A\pi} + c_2 M_{A\pi}\big]\;, \\
{\mathcal A}\big(\bar B^0\to K^-[ D^*(2010)^+\to] D^0\pi^+\big)&=&\frac{G_F}{\sqrt2}V_{cb}V^*_{us}
\big[\big(\frac{c_1}{3}+c_2\big)F_{TD^*} +c_1M_{TD^*}\big]\;,\\ 
{\mathcal A}\big(B^-\to \pi^-[ D^*(2007)^0\to]D^+\pi^-\big)&=&\frac{G_F}{\sqrt2}V_{cb}V^*_{ud}
\big[\big(\frac{c_1}{3}+c_2\big)F_{TD^*} +c_1M_{TD^*} +\big(c_1+\frac{c_2}{3}\big) F_{T\pi} + c_2 M_{T\pi}\big]\;, \\
{\mathcal A}\big(B^-\to K^-[ D^*(2007)^0\to]D^+\pi^-\big)&=&\frac{G_F}{\sqrt2}V_{cb}V^*_{us}
\big[\big(\frac{c_1}{3}+c_2\big)F_{TD^*} +c_1M_{TD^*} +\big(c_1+\frac{c_2}{3}\big) F_{TK} + c_2 M_{TK}\big]\;, \\
{\mathcal A}\big(B^-\to \pi^-[ D^*(2007)^0\to]D^0\pi^0\big)&=&\frac{G_F}{\sqrt2}V_{cb}V^*_{ud}
\big[\big(\frac{c_1}{3}+c_2\big)F_{TD^*} +c_1M_{TD^*} +\big(c_1+\frac{c_2}{3}\big) F_{T\pi} + c_2 M_{T\pi}\big]\;, \\
{\mathcal A}\big(B^-\to K^-[ D^*(2007)^0\to]D^0\pi^0\big)&=&\frac{G_F}{\sqrt2}V_{cb}V^*_{us}
\big[\big(\frac{c_1}{3}+c_2\big)F_{TD^*} +c_1M_{TD^*} +\big(c_1+\frac{c_2}{3}\big) F_{TK} + c_2 M_{TK}\big]\;, 
\end{eqnarray}
in which $G_F$ is the Fermi coupling constant, $V$'s are the CKM matrix elements, the Wilson coefficients $c_1$ and $c_2$ will
appear in convolutions in momentum fractions and impact parameters $b$.

The amplitudes from Fig.1 are written as
\begin{eqnarray}
F_{TD^*} &=& 8\pi C_F m^4_B f_{\pi(K)} (\eta-1)\int dx_B dz\int b_B db_B b db \phi_B(x_B,b_B)\phi_{D\pi}(z,b,s) \nonumber\\
&\times&\big\{\big[\sqrt{\eta}(1-2z)+z+1 \big] E^{(1)}_{a}(t^{(1)}_{e})h(x_B,z,b_B,b)
+\left(\eta+r_c \right) E^{(2)}_{a}(t^{(2)}_{e})h(z,x_B,b,b_B) \big\}, \\
M_{TD^*} &=& 32\pi C_F m^4_B/\sqrt{2N_c}(\eta-1) \int dx_B dz dx_3\int b_B db_B b_3 db_3\phi_B(x_B,b_B)
\phi_{D\pi}(z,b,s)\phi^A \nonumber\\
&\times& \big\{ \left[\eta\left(x_3-z-1\right)+z\sqrt{\eta}-x_B-x_3+1)\right]E_{b}(t^{(1)}_{b})h^{(1)}_{b}(x_i,b_i)\nonumber\\
&+&\left[x_3\left(\eta-1\right)+z\left(\sqrt{\eta}-1\right) +x_B \right]E_{b}(t^{(2)}_{b})h^{(2)}_{b}(x_i,b_i) \big\}\;, \\
F_{T\pi(K)}&=& 8\pi C_F m^4_B F_{D\pi}(s)  \int dx_B dx_3\int b_B db_B b_3 db_3 \phi_B(x_B,b_B)\nonumber\\
&\times &\big\{\big[\phi^A(\eta-1)[x_3(\eta-1)-1]
 +r_0[\phi^P(\eta-1)(2x_3-1)+\phi^T(2 x_3(\eta-1)+\eta+1)]\big]E^{(1)}_{c}(t^{(1)}_{i}) \nonumber\\
&\times&  h(x_B,x_3(1-\eta),b_B,b_3)-\left[2r_0\phi^P(\eta(1-x_B)-1)+\eta(1-\eta) x_B\phi^A\right]E^{(2)}_{c}(t^{(2)}_{i}) \nonumber\\
&\times&h(x_3,x_B(1-\eta),b_3,b_B)\big\}\;,\\
M_{T\pi(K)} &=& 32\pi C_F m^4_B/\sqrt{2N_c} \int dx_B dz dx_3\int b_B db_B b db\phi_B(x_B,b_B)\phi_{D\pi}(z,b,s) \nonumber\\
&\times&\big\{\big[  \phi^A(\eta-1)[\sqrt{\eta} r_c+\eta(x_B+z-1)-x_B-z+1]
+\eta r_0[\phi^P(x_B+z-x_3)+\phi^T(x_B+z+x_3-2)] \nonumber\\
&+& r_0 x_3(\phi^P-\phi^T)\big] E_{d}(t^{(1)}_{d})h^{(1)}_{d}(x_i,b_i)+\big[(\eta-1)[(\eta-1)x_3+x_B-z]\phi^A+r_0x_3(\eta-1)(\phi^P+\phi^T)\nonumber\\
&+&r_0\eta(x_B-z)(\phi^T-\phi^P)\big]E_{d}(t^{(2)}_{d})h^{(2)}_{d}(x_i,b_i)\big\}\;,\\
F_{A\pi} &=& 8\pi C_F m^4_B f_B \int dz dx_3\int b db b_3 db_3 \phi_{D\pi}(z,b,s))\nonumber\\
&\times&\big\{ \big[(1-\eta)[((\eta-1)x_3+1)\phi^A +r_0 r_c\phi^P]
+r_0r_c(\eta+1)\phi^T\big]E^{(1)}_{e}(t^{(1)}_{a})h_{a}(z,x_3(1-\eta),b,b_3)\nonumber\\
&+& \left[(\eta-1)z\phi^A-2r_0 \sqrt{\eta}(\eta+z-1)\phi^P\right]
E^{(2)}_{e}(t^{(2)}_{a})h_{a}(x_3,z(1-\eta),b_3,b) \big\},\\
M_{A\pi} &=& 32\pi C_F m^4_B/\sqrt{2N_c} \int dx_B dz dx_3\int b_B db_B b_3 db_3\phi_B(x_B,b_B)\phi_{D\pi}(z,b,s)  \nonumber\\
&\times& \big\{ \big[ (\eta-1) [(\eta-1)(x_B+z)-\eta]\phi^A
+r_0 \sqrt {\eta} [(x_3+\eta+1-x_3 \eta)\phi^T+(1-\eta)(x_3-1)\phi^P] \nonumber\\
&+ &r_0\sqrt {\eta} (z+x_B)(\phi^P-\phi^T) \big] E_{f}(t^{(1)}_{f})h^{(1)}_{f}(x_i,b_i) +\big[ (\eta-1) [\eta(x_B+x_3-z-1)-x_3+1] \phi^A \nonumber\\
&+&r_0 \sqrt {\eta} [ (1-\eta)(x_3-1)(\phi^P-\phi^T)+(z-x_B)(\phi^P+\phi^T)]   \big]E_{f}(t^{(2)}_{f})h^{(2)}_{f}(x_i,b_i) \big\},
\end{eqnarray}
The evolution factors in the above factorization formulas are given by
\begin{eqnarray}
E^{(1)}_a(t)&=&\alpha_s(t){\rm exp}[-S_B(t)-S_C(t)] S_t(z)\;, \quad ~E^{(2)}_a(t)=\alpha_s(t){\rm exp}[-S_B(t)-S_C(t)]  S_t(x_B)\;, \\
E_b(t) &=& \alpha_s(t){\rm exp}[-S_B(t)-S_C(t)-S_P(t)]|_{b=b_B}\;,\\
E^{(1)}_c(t)&=&\alpha_s(t){\rm exp}[-S_B(t)-S_P(t)] S_t(x_3)\;, \quad E^{(2)}_c(t)=\alpha_s(t){\rm exp}[-S_B(t)-S_P(t)]  S_t(x_B)\;, \\
E_d(t) &=& \alpha_s(t){\rm exp}[-S_B(t)-S_C(t)-S_P(t)]|_{b_3=b_B}\;,\\
E^{(1)}_e(t)&=&\alpha_s(t){\rm exp}[-S_C(t)-S_P(t)] S_t(x_3)\;, \quad E^{(2)}_e(t)=\alpha_s(t){\rm exp}[-S_C(t)-S_P(t)]  S_t(z)\;, \\
E_f(t) &=& \alpha_s(t){\rm exp}[-S_B(t)-S_C(t)-S_P(t)]|_{b_3=b}\;.
\end{eqnarray}
in which $S_{(B,C,P)}(t)$ are in the Appendix of~\cite{prd78-014018}, the hard functions $h, h_a, h^{(1,2)}_{(b,d,f)}$ and the 
hard scales $t^{(1,2)}_{(e,b,i,d,a,f)}$ have their explicit expressions in the Ref.~\cite{prd78-014018}. %We need to stress that, 
Because of the different definitions of the momenta for the initial and final states, the concerned expressions in~\cite{prd78-014018}
could be employed in this work only after the replacements $\{x_1\to x_B, b_1\to b_B, x_2\to z, b_2\to b, r^2\to\eta\}$.
The parameter $c$ in the Eq.~(A1) of~\cite{prd78-014018} is adopt to be 0.35 in this work according to the 
Refs.~\cite{prd65-014007,prd80-074024}.

%=============== Refs ===============%  


\begin{thebibliography}{199}
\bibitem{CKM-C}
N. Cabibbo, Phys. Rev. Lett. 10 (1963) 531.
\bibitem{CKM-KM}
M. Kobayashi, T. Maskawa, Prog. Theor. Phys. 49 (1973) 652.
%-----xxxxx CKM-gamma
\bibitem{prd67-096002}
R. Aleksan, T.C. Petersen, A. Soffer, Phys. Rev. D 67 (2003) 096002.
% "Measuring the Weak Phase gamma in Color Allowed B->DKpi Decays" C=63
\bibitem{prd79-051301}
T. Gershon, Phys. Rev. D 79 (2009) 051301.
% "On the Measurement of the Unitarity Triangle Angle gamma from B^0 -> DK^{*0} Decays" C=51
\bibitem{prd80-092002}
T. Gershon, M. Williams, Phys. Rev. D 80 (2009) 092002.
% "Prospects for the Measurement of the Unitarity Triangle Angle gamma from B0 ---> DK+ pi- Decays" C=42
\bibitem{prd81-014025}
T. Gershon, A. Poluektov, Phys. Rev. D 81 (2010) 014025.
% "Double Dalitz Plot Analysis of the Decay B0 ---> DK+ pi-, D ---> K0(S) pi+ pi-" C=15
\bibitem{prd97-056002}
D. Craik, T. Gershon, A. Poluektov, Phys. Rev. D 97 (2018) 056002.
% "Optimising sensitivity to $?$ with $B^0 \to DK^+?^-$, $D \to K_{\rm S}^0?^+?^-$ double Dalitz plot analysis" C=4
\bibitem{prd78-034023}
B. Aubert, et al., BaBar Collaboration, Phys. Rev. D 78 (2008) 034023.
% "Improved measurement of the CKM angle gamma in B^-+ -> D^(*) K^(*)-+ decays with a Dalitz plot analysis of
%% D decays to K_S^0 pi^+ pi^- and K_S^0 K^+ K^-"  C=168
%\bibitem{prd92-112005}
%R. Aaij {\it et al.} (LHCb collaboration), Phys. Rev. D {\bf 92,} 112005 (2015).
\bibitem{prd93-112018}
R. Aaij, et al., LHCb collaboration, Phys. Rev. D 93 (2016) 112018, Erratum: Phys. Rev. D 94 (2016) 079902.
\bibitem{jhep1612-087}
R. Aaij, et al., LHCb collaboration, J. High Energy Phys. 12 (2016) 087.
%-----xxxxx CKM-gamma
%-----xxxxx CKM-beta
\bibitem{plb425-375}
J. Charles, A. Le Yaouanc, L. Oliver, O. P\`{e}ne, J.C. Raynal, Phys. Lett. B 425 (1998) 375, Erratum: Phys. Lett. B 433 (1998) 441.
% "B0 (d)(t) ---> time Dependent dalitz plots, CP violating angles 2 Beta, 2 Beta + gamma, and discrete ambiguities" C=96
\bibitem{jpg36-025006}
T. Latham, T. Gershon, J. Phys. G 36 (2009) 025006.
\bibitem{jhep1803-195}
A. Bondar, A. Kuzmin, V. Vorobyev, J. High Energy Phys. 03 (2018) 195.
% "A method for model-independent measurement of the CKM angle $?$ via time-dependent analysis
%% of the $B^0\to D?^+?^-$, $D\to K_S^0?^+?^-$ decays" C=1
%-----xxxxx CKM-beta

%-----xxxxx DATA
\bibitem{prd69-112002}
K. Abe, et al., Belle Collaboration, Phys. Rev. D 69 (2004) 112002.
% "Study of B- => D**0 pi-(D**0 => D(*)+ pi-) decays" C=338
\bibitem{prd76-012006}
A. Kuzmin, et al., Belle Collaboration, Phys. Rev. D 76 (2007) 012006.
\bibitem{prd79-112004}
B. Aubert, et al., BaBar Collaboration, Phys. Rev. D 79 (2009) 112004.
% "Dalitz Plot Analysis of B- ---> D+ pi- pi-" C=60
\bibitem{prd91-092002}
R. Aaij, et al., LHCb collaboration, Phys. Rev. D 91 (2015) 092002, Erratum: Phys. Rev. D 93 (2016) 119901.
% "First observation and amplitude analysis of the $B^{-}\to D^{+}K^{-}?^{-}$ decay" C=41
\bibitem{prd92-012012}
R. Aaij, et al., LHCb collaboration, Phys. Rev. D 92 (2015) 012012.
% " Amplitude analysis of $B^0 \rightarrow \bar{D}^0 K^+ \pi^-$ decays" C=22
\bibitem{prd92-032002}
R. Aaij, et al., LHCb collaboration, Phys. Rev. D 92 (2015) 032002.
% "Dalitz plot analysis of $B^0 \to \overline{D}^0 ?^+?^-$ decays" C=55
\bibitem{prd93-051101}
R. Aaij, et al., LHCb collaboration, Phys. Rev. D 93 (2016) 051101.
\bibitem{prd94-072001}
R. Aaij, et al., LHCb collaboration, Phys. Rev. D 94 (2016) 072001.
% "Amplitude analysis of $B^{-} \to D^{+} pi^{-} pi^{-}$ decays" C=28
%-----xxxxx DATA

\bibitem{Goldhaber}
G. Goldhaber, et al., Phys. Lett. B 69 (1977) 503.
\bibitem{Nguyen1977}
H.K. Nguyen, et al., Phys. Rev. Lett. 39 (1977) 262.
\bibitem{Peruzzi}
I. Peruzzi, et al., Phys. Rev. Lett. 39 (1977) 1301.

\bibitem{npb591-313}
M. Beneke, G. Buchalla, M. Neubert, C.T. Sachrajda, Nucl. Phys. B 591 (2000) 313.
% "QCD factorization for exclusive, non-leptonic B meson decays: General arguments and the case of heavy-light final states" C=1153
\bibitem{jhep1609-112}
T. Huber, S. Kr\"ankl, X.Q. Li, J. High Energy Phys. 09 (2016) 112.
% "Two-body non-leptonic heavy-to-heavy decays at NNLO in QCD factorization" C=10

%-----xxxxx QCDF
\bibitem{zpc29-637}
M. Wirbel, B. Stech, M. Bauer, Z. Phys. C 29 (1985) 637.
% "Exclusive Semileptonic Decays of Heavy Mesons" C=1707
\bibitem{zpc34-103}
M. Bauer, B. Stech, M. Wirbel, Z. Phys. C 34 (1987) 103.
% "Exclusive Nonleptonic Decays of D, D(s), and B Mesons" C=1800
\bibitem{plb318-549}
A. Deandrea, N. Di Bartolomeo, R. Gatto, G. Nardulli, Phys. Lett. B 318 (1993) 549.
% "Two Body Non Leptonic Decays of B and B_s Mesons" C=218
\bibitem{ijmpa24-5845}
K. Azizi, R. Khosravi, F. Falahati, Int. J. Mod. Phys. A 24 (2009) 5845.
% "Analyzing of the B(q) ---> D(q)(D*(q)) P and B(q) ---> D(q) (D*(q)) V decays within the factorization approach in QCD" C=15
\bibitem{epjc76-523}
Q. Chang, L.X. Chen, Y.Y. Zhang, J.F. Sun, Y.L. Yang, Eur. Phys. J. C 76 (2016) 523.
% "$\bar{B}_{d,s} \to D^{*}_{d,s} V$ and $\bar{B}_{d,s}^* \to D_{d,s} V$ decays within QCD Factorization and Possible Puzzles" C=7
%-----xxxxx QCDF
\bibitem{prd75-074021}
C.W. Chiang, E. Senaha, Phys. Rev. D 75 (2007) 074021.
% "Update analysis of two-body charmed $B$ meson decays" C=14
\bibitem{prd92-094016}
S.H. Zhou, et al., Phys. Rev. D 92 (2015) 094016.
% "Analysis of Two-body Charmed $B$ Meson Decays in Factorization-Assisted Topological-Amplitude Approach" C=12
\bibitem{prd67-074013}
C.W. Chiang, J.L. Rosner, Phys. Rev. D 67 (2003) 074013.
% "Final-State Phases in $B \to D\pi, D^*\pi$, and $D\rho$ Decays" C=38

%-----xxxxx PQCD
\bibitem{plb504-6}
Y.Y. Keum, H.n. Li, A.I. Sanda, Phys. Lett. B 504 (2001) 6.
% "Fat penguins and imaginary penguins in perturbative QCD" C=648
\bibitem{prd63-054008}
Y.Y. Keum, H.n. Li, A.I. Sanda, Phys. Rev. D 63 (2001) 054008.
% "Penguin Enhancement and $B\to K?$ decays in perturbative QCD" C=743
\bibitem{prd63-074009}
C.D. L\"u, K. Ukai, M.Z. Yang, Phys. Rev. D 63 (2001) 074009.
% C=432
\bibitem{ppnp51-85}
H.n. Li, Prog. Part. Nucl. Phys. 51 (2003) 85.
% "QCD aspects of exclusive B meson decays" C=98
\bibitem{prd67-054028}
T. Kurimoto, H.n. Li, A.I. Sanda, Phys. Rev. D 67 (2003) 054028.
%  "$B \to D^{(*)}$ form factors in perturbative QCD" C=85
\bibitem{prd69-094018}
Y.Y. Keum, T. Kurimoto, H.n. Li, C.D. L\"u, A.I. Sanda, Phys. Rev. D 69 (2004) 094018.
% "Nonfactorizable contributions to $B \to D^{(*)} M$ decays" C=103
\bibitem{prd68-097502}
C.D. L\"{u}, Phys. Rev. D 68 (2003) 097502.
% "Study of color suppressed modes $B^0 \to \bar D^{(*)0} ?^{(\prime)} $" C=41
\bibitem{prd78-014018}
R.H. Li, C.D. L\"u, H. Zou, Phys. Rev. D 78 (2008) 014018.
%"The $B(B_s)\to D_{(s)} P$, $D_{(s)} V$, $D_{(s)}^{*}P$ and $D_{(s)}^{*}V$ decays in the perturbative QCD approach"
\bibitem{prd95-016011}
Z.T. Zou, Y. Li, X. Liu, Phys. Rev. D 95 (2017) 016011.
% "Two-body charmed B(Bs) decays involving a light scalar meson" C=2
%-----xxxxx PQCD

\bibitem{dalitz}
R.H. Dalitz, Philos. Mag. 44 (1953) 1068.

\bibitem{1806-09853}
A. Le Yaouanc, J.P. Leroy, P. Roudeau, arXiv:1806.09853.
% "Large off-shell effects in the $\bar{D}^*$ contribution to $B\to\bar{D}pipi$ and $B \to \bar{D}pi\bar{\ell} ?_{\ell}$ decays.}}" C=1

%%======symmetries=======
\bibitem{plb564-90}
M. Gronau, J.L. Rosner, Phys. Lett. B 564 (2003) 90.
% I-spin, U-spin, and penguin dominance in B --> K K anti-K
\bibitem{prd72-075013}
G. Engelhard, Y. Nir, G. Raz, Phys. Rev. D 72 (2005) 075013.
% SU(3) relations and the CP asymmetry in B ---> K(S) K(S) K(S)
\bibitem{prd72-094031}
M. Gronau, J.L. Rosner, Phys. Rev. D 72 (2005) 094031.
% Symmetry relations in charmless B ---> PPP decays
\bibitem{prd84-056002}
M. Imbeault, D. London, Phys. Rev. D 84 (2011) 056002.
% SU(3) Breaking in Charmless B Decays
\bibitem{plb727-136} %## U-spin breaking
M. Gronau, Phys. Lett. B 727 (2013) 136.
% U-spin breaking in CP asymmetries in B decays
\bibitem{plb726-337}
B. Bhattacharya, M. Gronau, J.L. Rosner, Phys. Lett. B 726 (2013) 337.
% CP asymmetries in three-body B+- decays to charged pions and kaons
\bibitem{prd89-074043}
B. Bhattacharya, et al., Phys. Rev. D 89 (2014) 074043.
% Charmless B-->PPP decays: The fully-symmetric final state
\bibitem{plb728-579}
D. Xu, G.N. Li, X.G. He, Phys. Lett. B 728 (2014) 579.
% U-spin analysis of CP violation in $B^-$ decays into three charged light pseudoscalar mesons
\bibitem{ijmpa29-1450011}
D. Xu, G.N. Li, X.G. He, Int. J. Mod. Phys. A 29 (2014) 1450011.
% Large SU3 breaking effects and CPV in B+ decays into three charged SU3 octet pseudoscalar mesons
\bibitem{prd91-014029}
X.G. He, G.N. Li, D. Xu, Phys. Rev. D 91 (2015) 014029.
% SU(3) and isospin breaking effects on B2PPP amplitudes
%===== A. Furman ====%
\bibitem{plb622-207}
A. Furman, R. Kami\'nski, L. Le\'sniak, B. Loiseau, Phys. Lett. B 622 (2005) 207.
% Long-distance effects and final state interactions in B-> pi pi K and B-> K anti-K K decays
\bibitem{prd74-114009}
B. El-Bennich, et al., Phys. Rev. D 74 (2006) 114009.
% Interference between f0(980) and rho(770) resonances in B --> pion-pion-kaon decays
\bibitem{prd79-094005}
B. El-Bennich, et al., Phys. Rev. D 79 (2009) 094005, Erratum: Phys. Rev. D 83 (2011) 039903.
% CP violation and kaon-pion interactions in B ---> K pi+ pi- decays
\bibitem{prd81-094033}
O. Leitner, J.-P. Dedonder, B. Loiseau, R. Kami\'nski, Phys. Rev. D 81 (2010) 094033, Erratum: Phys. Rev. D 82 (2010) 119906.
% K* resonance effects on direct CP violation in B -> pi pi K
\bibitem{appb42-2013}
J.-P. Dedonder, et al., Acta Phys. Polon. B 42 (2011) 2013.
% S-, P- and D-wave final state interactions and CP violation in B+- --> pi+- pi-+ pi+- decays
%===== Cheng Hai-Yang ====
\bibitem{prd66-054015}
H.Y. Cheng, K.C. Yang, Phys. Rev. D 66 (2002) 054015.
% Nonresonant three-body decays of D and B mesons
\bibitem{prd72-094003}
H.Y. Cheng, C.K. Chua, A. Soni, Phys. Rev. D 72 (2005) 094003.
% CP-violating asymmetries in B0 decays to K+ K- K0(S(L)) and K0(S) K0(S) K0(S(L))
\bibitem{prd76-094006}
H.Y. Cheng, C.K. Chua, A. Soni, Phys. Rev. D 76 (2007) 094006.
% Charmless three-body decays of B mesons
\bibitem{prd88-114014}
H.Y. Cheng, C.K. Chua, Phys. Rev. D 88 (2013) 114014.
% Branching Fractions and Direct CP Violation in Charmless Three-body Decays of B Mesons
\bibitem{prd89-074025}
H.Y. Cheng, C.K. Chua, Phys. Rev. D 89 (2014) 074025.
% Charmless three-body decays of Bs mesons
\bibitem{prd89-094007}
Y. Li, Phys. Rev. D 89 (2014) 094007.
% Comprehensive Study of B->KKpi Decays in the Factorization Approach
%===================
\bibitem{epjc75-536}
C. Wang, Z.H. Zhang, Z.Y. Wang, X.H. Guo, Eur. Phys. J. C 75 (2015) 536.
\bibitem{prd87-076007}
Z.H. Zhang, X.H. Guo, Y.D. Yang, Phys. Rev. D 87 (2013) 076007.
% CP violation in $B\to 3\pi$ in the region with low invariant mass of one $\pi^+\pi^-$ pair
\bibitem{epjc78-845}
J.J. Qi, Z.Y. Wang, Z.H. Zhang, J. Xu, X.H. Guo, Eur. Phys. J. C 78 (2018) 845.

\bibitem{npb899-247}
S. Kr\"ankl, T. Mannel, J. Virto, Nucl. Phys. B 899 (2015) 247.
% Three-Body Non-Leptonic B Decays and QCD Factorization
\bibitem{prd96-113003}
D. Boito, et al., Phys. Rev. D 96 (2017) 113003.
% "Parametrizations of three-body hadronic B- and D-decay amplitudes
%% in terms of analytic and unitary meson-meson form factors"  C=3

\bibitem{plb561-258}
C.H. Chen, H.n. Li, Phys. Lett. B 561 (2003) 258.
\bibitem{prd70-054006}
C.H. Chen, H.n. Li, Phys. Rev. D 70 (2004) 054006.
\bibitem{prd89-074031}
W.F. Wang, H.C. Hu, H.n. Li, C.D. L\"u, Phys. Rev. D 89 (2014) 074031.
% "Direct CP asymmetries of three-body B decays in perturbative QCD" C=30
\bibitem{prd91-094024}
W.F. Wang, H.n. Li, W. Wang, C.D. L\"u, Phys. Rev. D 91 (2015) 094024.
% "$S$-wave resonance contributions to the $B^0_{(s)}\to J/??^+?^-$ and $B_s\to?^+?^-?^+?^-$ decays" C=36
\bibitem{prd97-034033}
C. Wang, J.B. Liu, H.n. Li, C.D. L\"u, Phys. Rev. D 97 (2018) 034033.
% "Three-body decays $B \to ?(?) K ?$ in perturbative QCD approach"
\bibitem{1803-02656}
N. Wang, Q. Chang, Y.L. Yang, J.F. Sun, arXiv:1803.02656.
\bibitem{1810-12507}
Z.R. Liang, F.B. Duan, X.Q. Yu, arXiv:1810.12507.

\bibitem{1512-09284}
I. Bediaga, P.C. Magalh\~aes, arXiv:1512.09284.
% Final state interaction on $B^+\to \pi^-\pi^+\pi^+$
%\bibitem{plb780-357}
%I.~Bediaga, T.~Frederico, and P.~C.~Magalh\~aes, Phys. Lett. B {\bf 780,} 357 (2018).
%``Charm Penguin in $B^\pm \to K^\pm K^+ K^-$: partonic and hadronic loops,'' C=4
%\bibitem{epjc77-561}
%J. Charles, S. Descotes-Genon, J. Ocariz, and A. P\'erez P\'erez, Eur. Phys. J. C {\bf 77,} 561 (2017).

\bibitem{plb763-29}
W.F. Wang, H.n. Li, Phys. Lett. B 763 (2016) 29.

\bibitem{1605-03889}
J.H.A. Nogueira, et al., arXiv:1605.03889.
% "Summary of the 2015 LHCb workshop on multi-body decays of D and B mesons" C=12

\bibitem{paps-li-ya}
Y. Li, A.J. Ma, W.F. Wang, Z.J. Xiao, Phys. Rev. D 95 (2017) 056008; Phys. Rev. D 96 (2017) 036014.
\bibitem{paps-li-ya-II}
Y. Li, W.F. Wang, A.J. Ma, Z.J. Xiao, Eur. Phys. J. C  79 (2019) 37. %%arXiv:1809.09816[hep-ph].  (2019) 79:37
\bibitem{paps-ma-aj}
A.J. Ma, Y. Li, W.F. Wang, Z.J. Xiao, Nucl. Phys. B 923 (2017) 54; Phys. Rev. D 96 (2017) 093011.
\bibitem{1809-02943}
W.F. Wang, Phys. Lett. B 788 (2019) 468.
% "Resonant state $D_0^\ast(2400)$ in the quasi-two-body $B$ meson decays"

\bibitem{prd86-114025}
W.F. Wang, Z.J. Xiao, Phys. Rev. D 86 (2012) 114025.
% "The semileptonic decays $B/B_s \to (?, K)(l^+l^-,l?,?\bar?)$ in the pQCD approach beyond the leading-order" C=57

\bibitem{epjc78-76}
Y. Zhang, et al., Eur. Phys. J. C 78 (2018) 76.
% "Uncertainties of the $B\to D$ transition form factor from the $D$-Meson Leading-Twist Distribution Amplitude" C=1
\bibitem{1807-03453}
T. Zhong, Y. Zhang, X.G. Wu, H.B. Fu, T. Huang, Eur. Phys. J. C 78 (2018) 937.
%"The Ratio $\mathcal{R}(D)$ and the $D$-meson Distribution Amplitude" C=0

%=============== f_D*
\bibitem{ijmpa30-1550116}  %%OK%%
S. Narison, Int. J. Mod. Phys. A 30 (2015) 1550116.
% "Improved f_{D*_(s)}, f_{B*_(s)} and f_{B_c} from QCD Laplace sum rules" C=20
\bibitem{nppp270-143}
S. Narison, Nucl. Part. Phys. Proc. 270-272 (2016) 143.
% "Decay Constants of Heavy-Light Mesons from QCD" C=13
\bibitem{epjc75-427}
Z.G. Wang, Eur. Phys. J. C 75 (2015) 427 and the references.
%``Analysis of the masses and decay constants of the heavy-light mesons with QCD sum rules'' C=28
\bibitem{prd96-034524}
V. Lubicz, A. Melis, S. Simula, ETM Collaboration, Phys. Rev. D 96 (2017) 034524.
% "Masses and decay constants of D(s)* and B(s)* mesons with Nf = 2 + 1 + 1 twisted mass fermions" C=9
\bibitem{1810-00296}
Q. Chang, X.N. Li, X.Q. Li, F. Su, Y.D. Yang, arXiv:1810.00296 and the references.
% " Self-consistency and covariance of light-front quark models: testing via $P$, $V$ and $A$ meson decay constants,
% and $P\to P$ weak transition form factors" C=0
%=============== f_D*

\bibitem{BW-factor1952}
J. Blatt, V. Weisskopf, {\it Theoretical Nuclear Physics,} (John Wiley \& Sons, New York, 1952).

\bibitem{prd90-072003}
R. Aaij, et al., LHCb collaboration, Phys. Rev. D 90 (2014) 072003.
%"Dalitz plot analysis of $B_s^0 \rightarrow \bar{D}^0 K^- ?^+$ decays" arXiv:1407.7712

\bibitem{prd65-032003}
A. Anastassov, et al., CLEO Collaboration, Phys. Rev. D 65 (2002) 032003.
%``First measurement of Gamma(D*+) and precision measurement of m(D*+) - m(D0)'' C=190

\bibitem{prd88-052003}
J.P. Lees, et al., BaBar Collaboration, Phys. Rev. D 88 (2013) 052003, Erratum: Phys. Rev. D 88 (2013) 079902.
% "Measurement of the D*(2010)+ natural line width and the D*(2010)+ - D0 mass difference" C=28
\bibitem{prl111-111801}
J.P. Lees, et al., BaBar Collaboration, Phys. Rev. Lett. 111 (2013) 111801, Erratum: Phys. Rev. Lett. 111 (2013) 169902.
%``Measurement of the $D*(2010)^+$ meson width and the $D*(2010)^+ - D^0$ mass difference,'' C=33

\bibitem{PDG-2018}
M. Tanabashi, et al., Particle Data Group, Phys. Rev. D 98 (2018) 030001.

\bibitem{prd98-074512} %1712-09262
A. Bazavov, et al., FNAL and MILC Collaborations, Phys. Rev. D 98 (2018) 074512.
% "B and D meson leptonic decay constants from four-flavor lattice QCD" C=16

\bibitem{prd90-094018}
W.F. Wang, X. Yu, C.D. L\"u, Z.J. Xiao, Phys. Rev. D 90 (2014) 094018.
% "Semileptonic decays $B_c^+\to D^{(*)}_{(s)}(l^+?,l^+l^-,?\bar?)$ in the perturbative QCD approach" C=10

\bibitem{prl96-011803}
B. Aubert, et al., BaBar Collaboration, Phys. Rev. Lett. 96 (2006) 011803.
% "Measurement of Branching Fractions and Resonance Contributions for B0->D0barK+pi- and Search for B0->D0K+pi- Decays" C=24
\bibitem{prl87-111801}
K. Abe, et al., Belle Collaboration, Phys. Rev. Lett. 87 (2001) 111801.
% "Observation of Cabibbo suppressed $B \to D^{(*)}K^-$ decays at Belle" C=34

\bibitem{plb777-16}
R. Aaij, et al., LHCb collaboration, Phys. Lett. B 777 (2018)16.
% "Measurement of $CP$ observables in $B^\pm \to D^{(*)} K^\pm$ and $B^\pm \to D^{(*)} ?^\pm$ decays" C=17

\bibitem{prd71-031102}
B. Aubert, et al., BaBar Collaboration, Phys. Rev. D 71 (2005) 031102(R).
% "Measurement of the Ratio BR(B- --> D*0 K-)/BR(B- --> D*0 pi-) and of the CP Asymmetry of B- --> D*0(CP+) K- Decays" C=27

\bibitem{prd78-014029}
X.H. Zhong, Q. Zhao, Phys. Rev. D 78 (2008) 014029.
%"Strong decays of heavy-light mesons in a chiral quark model" %Cited-77

\bibitem{jhep1404-177}
C.Y. Cheung, C.W. Hwang, J. High Energy Phys. 04 (2014) 177.
% "Strong and radiative decays of heavy mesons in a covariant model" C=17

\bibitem{prd88-034034}
J.L. Rosner, Phys. Rev. D 88 (2013) 034034.
% "Hadronic and radiative $D^*$ decays" C=6

\bibitem{plb721-94}
D. Be\v{c}irevi\'c, F. Sanfilippo, Phys. Lett. B 721 (2013) 94.
 % "Theoretical estimate of the $D^* \to D?$ decay rate" C=43

\bibitem{prd97-012005}
Y. Kato, et al., Belle Collaboration, Phys. Rev. D 97 (2018) 012005.
% "Measurements of the absolute branching fractions of $B^{+} \to X_{c\bar{c}} K^{+}$ and $B^{+} \to \bar{D}^{(\ast) 0} ?^{+} $
%  at Belle" C=4


\bibitem{prd65-014007}
T. Kurimoto, H.n. Li, A.I. Sanda, Phys. Rev. D 65~(2001)  014007.
% "Leading-power contributions to $B\to pi,\rho$ transition form factors"  C=228

\bibitem{prd80-074024}
H.n. Li, S. Mishima, Phys. Rev. D  80 (2009)  074024. 

%%%% ---ZZZZZZZ----------
%Y. Amhis {\it et al.} (HFLAV Collaboration), Eur. Phys. J. C {\bf 77,} 895 (2017).
% "Averages of $b$-hadron, $c$-hadron, and $?$-lepton properties as of summer 2016" C=454
\end{thebibliography}
\end{document}